\documentclass[superscriptaddress,preprint,floatfix,amsmath,aps]{revtex4-1}
\usepackage{amsmath,amssymb,latexsym}
\usepackage{graphicx}
\usepackage{hyperref}
\usepackage{subfigure}
\usepackage{appendix}

\begin{document}

\title{Coexistence of nontrivial topological properties and strong ferromagnetic fluctuations in quasi-one-dimensional $A_2$Cr$_3$As$_3$}

 \affiliation{Department of Physics, Zhejiang University, Hangzhou 310027, China}
 \affiliation{Condensed Matter Group,
  Department of Physics, Hangzhou Normal University, Hangzhou 311121, China}
 \affiliation{Shanghai Branch, National Laboratory for Physical Sciences at Microscale and Department of Modern Physics, University
  of Science and Technology of China, Shanghai 201315, China} 
 \affiliation{Westlake University, Hangzhou, Zhejiang,China}
 \affiliation{Department of Physics, Fudan University, Shanghai,China}
 \affiliation{Collaborative Innovation Center of Advanced Microstructures, Nanjing University, Nanjing 210093, China}
 \affiliation{CAS Key Laboratory of Functional Materials and Devices for Special Environments, Xinjiang Technical Institute of Physics and Chemistry, CAS, Xinjiang Key Laboratory of Electronic Information Materials and Devices, 40-1 South Beijing Road, Urumqi 830011, China}

\author{Chenchao Xu}
 \affiliation{Department of Physics, Zhejiang University, Hangzhou 310027, China}

\author{Ninghua Wu}
 \affiliation{Department of Physics, Zhejiang University, Hangzhou 310027, China}

\author{Guo-Xiang Zhi}
 \affiliation{Department of Physics, Zhejiang University, Hangzhou 310027, China}

\author{Bing-Hua Lei}
 \affiliation{CAS Key Laboratory of Functional Materials and Devices for Special Environments, Xinjiang Technical Institute of Physics and Chemistry, CAS, Xinjiang Key Laboratory of Electronic Information Materials and Devices, 40-1 South Beijing Road, Urumqi 830011, China}

\author{Xu Duan}
 \affiliation{Condensed Matter Group,
  Department of Physics, Hangzhou Normal University, Hangzhou 311121, China}
 \affiliation{Westlake University, Hangzhou, Zhejiang,China}
 \affiliation{Department of Physics, Fudan University, Shanghai,China}

\author{Fanlong Ning}
 \affiliation{Department of Physics, Zhejiang University, Hangzhou 310027, China}
 \affiliation{Collaborative Innovation Center of Advanced Microstructures, Nanjing University, Nanjing 210093, China}

\author{Chao Cao}
 \email[E-mail address: ]{ccao@hznu.edu.cn}
 \affiliation{Condensed Matter Group,
  Department of Physics, Hangzhou Normal University, Hangzhou 311121, China}
 \affiliation{Department of Physics, Zhejiang University, Hangzhou 310027, China}

\author{Qijin Chen}
 \email[E-mail address:] {qchen@zju.edu.cn}
 \affiliation{Department of Physics, Zhejiang University, Hangzhou 310027, China}
 \affiliation{Shanghai Branch, National Laboratory for Physical Sciences at Microscale and Department of Modern Physics, University
  of Science and Technology of China, Shanghai 201315, China}

\date{\today}

\begin{abstract}
  Superconductivity in crystals without inversion symmetry has  received extensive attention due to its unconventional pairing and  possible nontrivial topological properties. Using first-principles  calculations, we systemically study the electronic structure of noncentrosymmetric superconductors $A_2$Cr$_3$As$_3$ ($A$=Na, K, Rb  and Cs). Topologically protected triply degenerate points connected by one-dimensional arcs appear along the $C_{3}$ axis, coexisting with strong ferromagnetic (FM) fluctuations in the non-superconducting state. Within random phase approximation, our calculations show that strong enhancements of spin fluctuations are present in K$_2$Cr$_3$As$_3$ and Rb$_2$Cr$_3$As$_3$, and are substantially reduced in Na$_2$Cr$_3$As$_3$ and Cs$_2$Cr$_3$As$_3$. Symmetry analysis of pairing gap $\Delta(\mathbf{k})$ and spin-orbit coupling $\mathbf{g}_{\mathbf{k}}$ suggest that the arc surface states may also exist in the superconducting state, giving rise to possible nontrivial topological properties.
\end{abstract}

\maketitle

\thispagestyle{empty}
\newpage

\section{Introduction}
Materials with nontrivial topological properties have been extensively studied over the last two decades. Initially, comprehensive attentions are paid to topological insulators (TIs) \cite{Qi_RMP,Bi2Se3_zhang,Fu_RMP}, which have an insulating gap in the bulk and metallic surface states at the boundary. Fully gapped superconductors with the topological protected gapless surface mode, a close analogy with TI, are regarded as promising candidates for hosting Majorana fermions. Since the discovery of Weyl, Dirac, nodal line semimetals and triply degenerate points topological metals \cite{Wan_Weyl,Weng_Weyl,Lv_TaAs,Na3Bi_XiDaiZhongFan,Cd3As2_XiDaiZhongFang,Na3Bi_liu,Cd3As2_Cava,Burkov_nodal_line,Weng_Cu3PdN,Weng_MTC,TP3_typeA,TP3_InAsSb}, it was shown that gapless systems can possess novel topology as well. Simultaneously, it is also possible for superconductors with nodes (e.g., CePt$_3$Si, UPt$_{3}$) to have topologically protected edge states, which are guaranteed by momentum dependent topological numbers\cite{Schnyder_NSC,UPt3_Timm,schnyder_nodal_topological_review}. Among these exotic superconductors, nodal noncentrosymmetric superconductors (NCSs) with topological stable nodes have fascinating properties, i.e. the zero-energy boundary modes \cite{PhysRevB.84.060504,PhysRevB.84.020501}. These zero-energy boundary modes are believed to be closely associated with the nodal gap structure via the so-called bulk-boundary correspondence\cite{matsuura2013protected}.

The Cr-based arsenides $A_2$Cr$_3$As$_3$ ($A$=Na, K, Rb and Cs) are of great interest in terms of low dimensionality, strong ferromagnetic (FM) fluctuations and noncentrosymmetric superconductivity. Their nodal, unconventional superconductivity was suggested by London penetration depth, NMR, $\mu$SR and specific-heat measurements \cite{pang2015evidence,pang2016penetration,NMR_K233,NMR_Rb233,muSR_K,muSR_Cs,shao_K_specific-heat,Rb_2015}. Besides, spin-orbit coupling (SOC) has a remarkable effect on $\beta$ and $\gamma$ bands in K$_2$Cr$_3$As$_3$ with a band spin-splitting much larger than the superconducting gap\cite{Jiang2015}. The coalescence of considerable SOC effect and strong FM fluctuations in $A_2$Cr$_3$As$_3$ NCSs is crucial to the superconducting pairing symmetry, leading to a predominant spin-triplet component, which is distinguished from the pairing in the isotropic channel of an usual $s$-wave superconductor. More recently, NMR experiments on $A_2$Cr$_3$As$_3$ suggest that the compounds in this family are possibly a solid-state analog of superfluid $^{3}$He, implying that the unconventional superconductor $A_2$Cr$_3$As$_3$ may host nontrivial topological properties\cite{Zheng_QCP}. In addition, the NMR measurements of Cs$_2$Cr$_3$As$_3$ were distinct from that of K$_2$Cr$_3$As$_3$ (Rb$_2$Cr$_3$As$_3$), which displayed suppression of FM fluctuations in the former\cite{NMR_Cs233}. Therefore, a systematic study of the SOC effect, the topological properties as well as the FM fluctuations, and a thorough comparison among the family is in need.

In this article, we report our latest first-principles results on the $A_2$Cr$_3$As$_3$ family. Our results show: 1. The variations in band structures and Fermi surfaces (FSs) due to alkaline element substitution do not show apparent systematic behavior; 2. The anti-symmetric spin-orbit coupling (ASOC) splitting is the largest in K$_2$Cr$_3$As$_3$, but its effect is most significant in Cs$_2$Cr$_3$As$_3$ and enhances its one dimensionality; 3. All compounds of this family host triply degenerate points (TPs) along $\Gamma$-$A$ and the surface states emerging from the TPs form one-dimensional (1D) Fermi arcs; 4. The magnetic susceptibility spectrum exhibits a strong peak of the spin susceptibilities at the $\Gamma$ point in K$_2$Cr$_3$As$_3$, followed by Rb$_2$Cr$_3$As$_3$, while in Na$_2$Cr$_3$As$_3$ and Cs$_2$Cr$_3$As$_3$ the enhancement of spin susceptibilities at the $\Gamma$ point is not obvious. Inclusion of dynamic self-energy due to spin fluctuation does not alter the topology of electronic spectrum, and thus the existence of TPs is robust against the dynamic spin fluctuation at RPA level. Finally, we discuss the possibility of the existence of topologically stable arc states in the superconducting phase.   

\section{Results}
\subsection{Electronic Structure and Topological Properties}
\begin{figure}
  \includegraphics[width=12 cm]{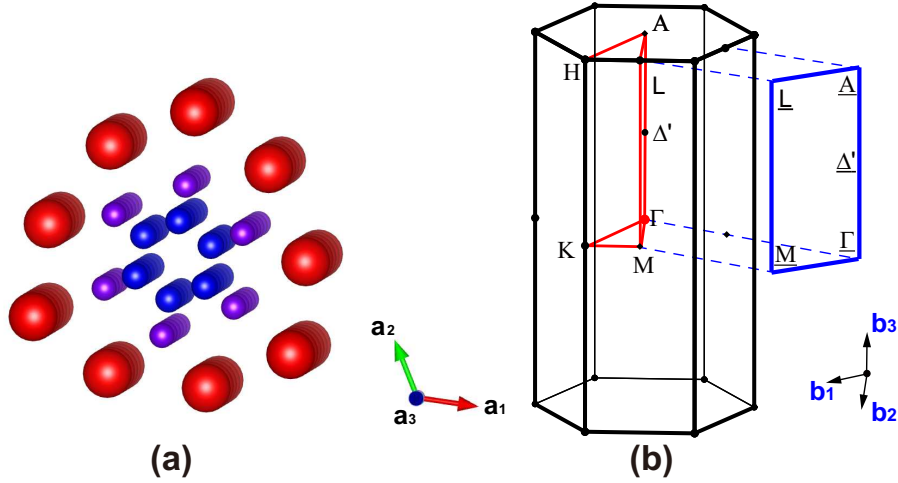}
  \caption{Geometry and Brillouine zone. (a) Crystal structure of $A_2$Cr$_3$As$_3$. The red, blue and purple are alkali, chromium and arsenic atoms, respectively. (b) Brillouine zone (BZ) and high symmetry points. The blue lines represent the [010] surface BZ .  }
  \label{fig:CRY_BZ}
\end{figure}

The crystal structure and Brillouin zone of K$_{2}$Cr$_{3}$As$_{3}$ are illustrated in Fig. \ref{fig:CRY_BZ} (a-b). For the compounds of this family, each primitive unit cell consists of [(Cr$_{6}$As$_{6}$)]$_{\infty}$ sub-nanotube along the $c$ axis forming a quasi-one-dimensional (Q1D) structure\cite{Na_2019,K_2015,Rb_2015,wang_K_Rb233,Cs_2015}. In contrast to $A$Cr$_{3}$As$_{3}$ ($A$=K,Rb) \cite{Bao_K133,ren_K133,tang_Rb_Cs133,ren_Rb133}, the $A^{+}$ ions around the [(Cr$_{6}$As$_{6}$)]$_{\infty}$ sub-nanotube break the inversion symmetry, rendering $A_2$Cr$_3$As$_3$ non-centrosymmetric (with symmetry $D_{3h}$, space group 187). From Na$_{2}$Cr$_{3}$As$_{3}$, K$_{2}$Cr$_{3}$As$_{3}$, Rb$_{2}$Cr$_{3}$As$_{3}$ to Cs$_{2}$Cr$_{3}$As$_{3}$, the lattice expands in-plane while the average Cr-Cr bond length barely changes, implying increased distances between the [(Cr$_{6}$As$_{6}$)]$_{\infty}$ tubes. 

\begin{figure}
  \includegraphics[width=15 cm]{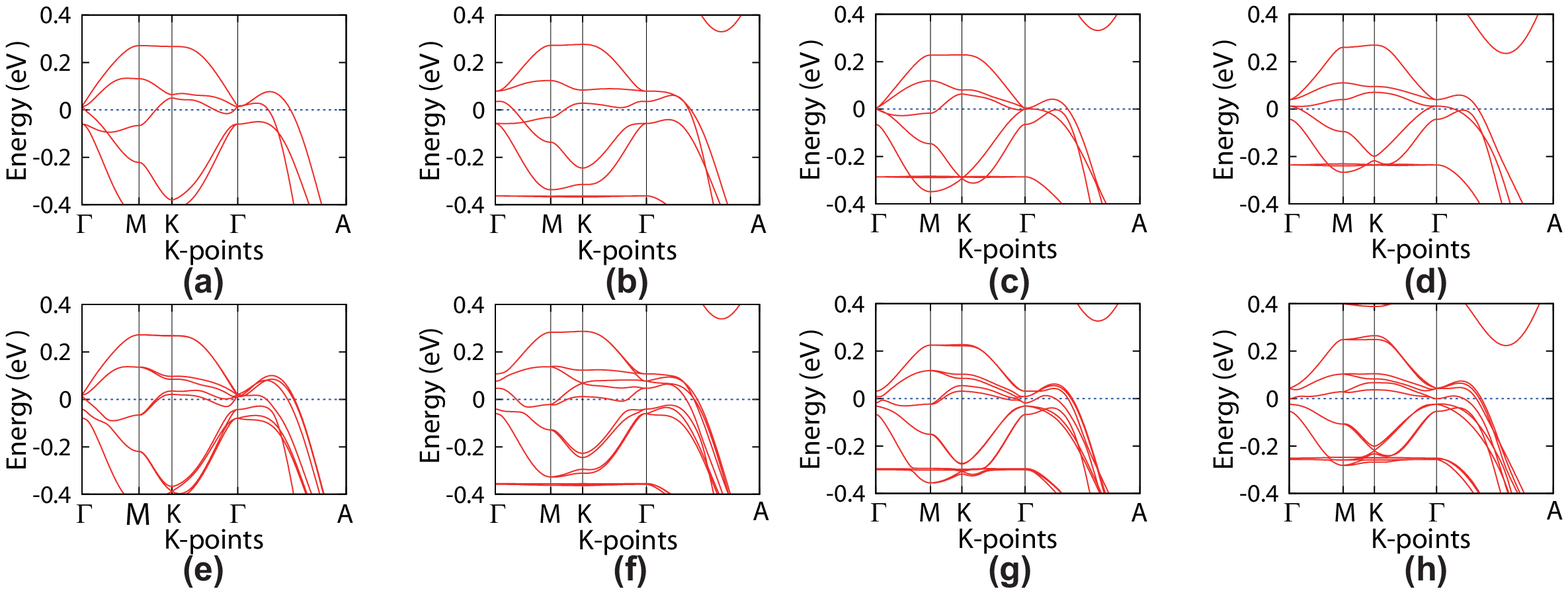}
  \caption{Electronic band structure of $A_2$Cr$_3$As$_3$. (a,e): $A$=Na; (b,f): $A$=K; (c,g): $A$=Rb; (d,h): $A$=Cs. The upper/lower panels are results without/with spin-orbit coupling.}
  \label{fig:band_structure}
\end{figure}

\begin{figure}
  \includegraphics[width=8 cm]{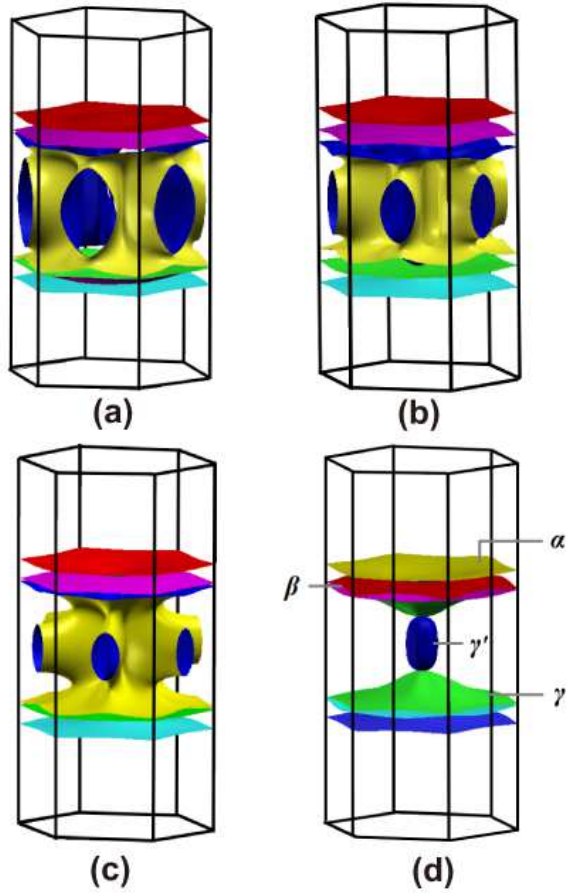}
  \caption{Fermi-surface sheets of $A_2$Cr$_3$As$_3$. (a) Na$_2$Cr$_3$As$_3$, (b) K$_2$Cr$_3$As$_3$, (c) Rb$_2$Cr$_3$As$_3$ and (d) Cs$_2$Cr$_3$As$_3$. In Na$_2$Cr$_3$As$_3$, K$_2$Cr$_3$As$_3$ and Rb$_2$Cr$_3$As$_3$, there are two Q1D FSs and one 3D FS, whereas in Cs$_2$Cr$_3$As$_3$ one more FS emerges. See also Supplementary Figure 1}
  \label{fig:FS}
\end{figure}

Despite this systematic structural variation, the band structures (Fig. \ref{fig:band_structure} (a-d)) of $A_2$Cr$_3$As$_3$ around the Fermi level resemble each other and do not exhibit apparent change except for Cs$_{2}$Cr$_{3}$As$_{3}$. In the absence of SOC, the $\alpha$ and $\beta$ bands cross the Fermi level along $\Gamma$-$A$, forming two Q1D FSs (please refer to Supplementary Figure 1 for details). In contrast, the $\gamma$ band forms one three-dimensional (3D) FS around the $\Gamma$ point (Fig. \ref{fig:FS} (a-c)) for Na$_{2}$Cr$_3$As$_3$ and Rb$_{2}$Cr$_3$As$_3$, similar to K$_{2}$Cr$_{3}$As$_{3}$ \cite{Jiang2015,xian2015magnetism}. For Cs$_{2}$Cr$_{3}$As$_{3}$, it is worth noting that the $\gamma$ band do not cross the Fermi level in the $k_{z}= 0$ plane. As a result, this band forms one deformed Q1D FS. Additionally, a fourth band ($\gamma'$) around the $\Gamma$ point emerges, creating a new 3D FS (Fig. \ref{fig:FS} (d)). Once the SOC effect is included, for all members in this family, the $\beta$ and $\gamma$ bands further split due to the ASOC effect.

Remarkably, the compounds of $A_2$Cr$_3$As$_3$ host TPs along $\Gamma$-$A$. In particular, the TPs in Na$_2$Cr$_3$As$_3$ (8 meV above the Fermi level $\epsilon_F$) and K$_2$Cr$_3$As$_3$ (0.6 meV below $\epsilon_F$) around the $\Gamma$ point are very close to the Fermi level. Considering the presence of time reversal symmetry $T$ and mirror symmetry $\sigma_{h}$ (orthogonal to the $C_{3}$ axis) in $A_2$Cr$_3$As$_3$, the TP fermions in current compounds belong to type A \cite{TP3_typeA}, which is accompanied by one Weyl nodal line instead of four in type B \cite{TP3_InAsSb,HgTe}. Including SOC, along the $C_{3}$ rotation axis, the singly degenerate band ($\Lambda_5$ or $\Lambda_6$ state) belongs to the 1D representation of $C_{3v}$ symmetry, while the doubly degenerate bands  ($\Lambda_4$ state) form the two-dimensional (2D) representation, and these TPs along $k_{z}$ are due to band inversion between the $\Lambda_5$/$\Lambda_6$ state and the $\Lambda_4$ state. Around the Fermi level band inversion occurs for $\alpha$, $\beta$ and $\gamma$ bands in Na$_2$Cr$_3$As$_3$ (Fig.~\ref{fig:Ss}(a)) as well as $\alpha$ and $\beta$ in K$_2$Cr$_3$As$_3$ (see Supplementary Figure 2), and these TPs are protected by the $C_{3v}$ symmetry. We compare the band structures of $A_2$Cr$_3$As$_3$ along $k_{z}$ (Supplementary Figure 2) as well as TPs and (010) surface states (see Supplementary Figure 3(a-c)). Due to the overwhelming bulk states, only the surface states in Na$_2$Cr$_3$As$_3$ can be clearly distinguished from the bulk band continuum, resulting in clear Fermi arc structures on the (010) surface (Fig.~\ref{fig:Ss}(b-c)). We also performed calculations using the modified Becke-Johnson (mBJ) potentials \cite{mBJ}, and obtained results similar to that of PBE for Na$_2$Cr$_3$As$_3$ (For details, see Methods below). In K$_2$Cr$_3$As$_3$, however, the $\Lambda_4$ state near $\epsilon_F$, which is lower than the $\Lambda_5$ and $\Lambda_6$ states in PBE calculations (see Supplementary Figure 2(b)), is now elevated higher around $\Gamma$ points, leading to extra band inversions between $\beta$ and $\gamma$ band and hence two new TPs (located at $\epsilon_F$+90 meV and $\epsilon_F$-22 meV), as shown in Fig.~\ref{fig:Ss}(d). Figure \ref{fig:Ss}(b-c) and (e-f) show the surface states in Na$_2$Cr$_3$As$_3$ (PBE results) and K$_2$Cr$_3$As$_3$ (mBJ results), respectively. For both of these two compounds, two surface states, $SS_{1}$ and $SS_{2}$, emerge from two TPs (TP$_{1}$ and TP$_{2}$) near the $\Gamma$ point, respectively, while the surface states from the TPs (TP$_{3}$ and TP$_{4}$ in Na$_2$Cr$_3$As$_3$, TP$_{3}$ in K$_2$Cr$_3$As$_3$) away from $\Gamma$ point are mixed with surrounding bulk states and cannot be easily distinguished. The iso-energy surface states at TP$_{1}$ in Na$_2$Cr$_3$As$_3$ ( $\epsilon_F+8$meV) and K$_2$Cr$_3$As$_3$ ($\epsilon_F+90$meV) are shown in Fig.~\ref{fig:Ss}(c) and (f). Akin to ZrTe family of compounds, the TPs (TP$_{1}$ in Na$_2$Cr$_3$As$_3$ and K$_2$Cr$_3$As$_3$) marked as blue (Fig.~\ref{fig:Ss}(c)) and red (Fig.~\ref{fig:Ss}(f)) dots are connected by double 1D Fermi arcs on the (010) surface. We note that for K$_2$Cr$_3$As$_3$, the number of TPs are different in PBE and mBJ calculations. Nevertheless, as long as the band-inversions between the $\Lambda_4$ and $\Lambda_5$ ($\Lambda_6$) exist, these intrinsic TPs will be protected by the extra global symmetry ($C_{3v}$).  In the next section, we will further discuss the dynamic correlation effect due to the spin fluctuation on these TPs.

\begin{figure}
  \includegraphics[width=15 cm]{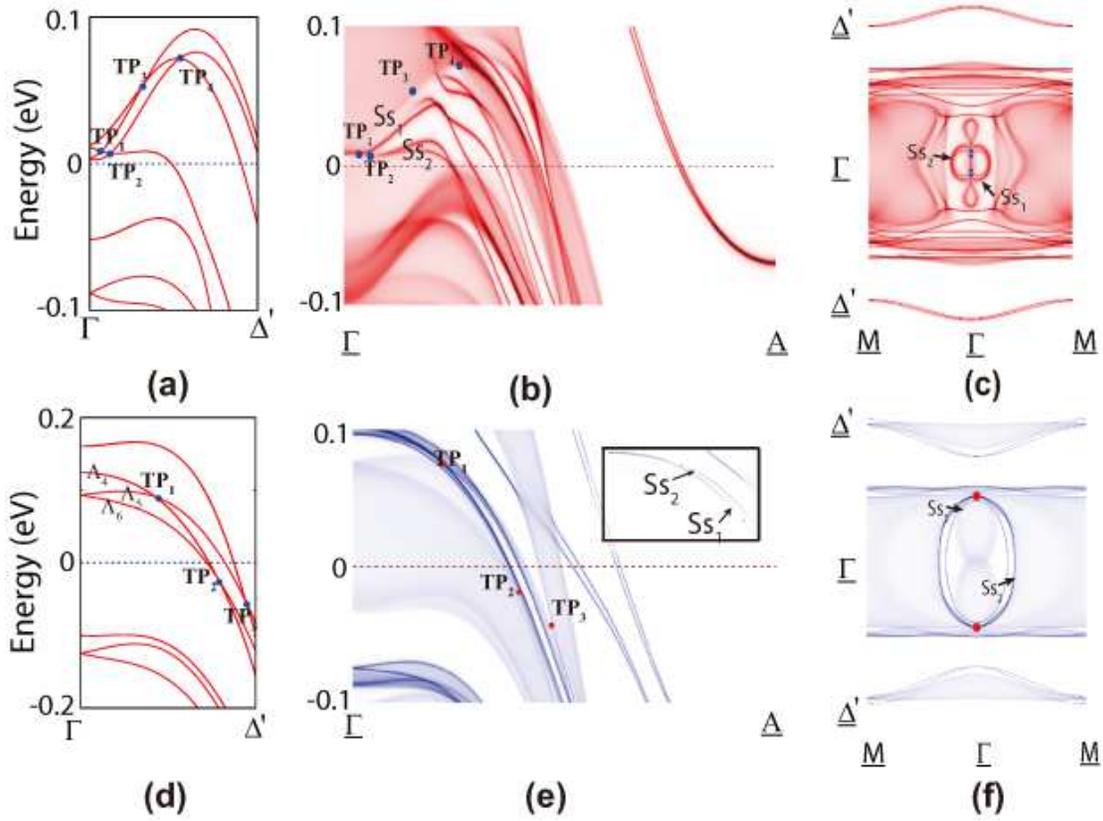}
  \caption{Electronic band structure and the (010) surface state along $k_{z}$. (a-c) Na$_2$Cr$_3$As$_3$ within PBE calculation. (d-e) K$_2$Cr$_3$As$_3$ within mBJ calculation. The triply degenerate points are marked as solid dots. The iso-energy surface state is located at $\epsilon_F$+8 meV for Na$_2$Cr$_3$As$_3$, and at $\epsilon_F$+90 meV for K$_2$Cr$_3$As$_3$. The inset of (f) shows surface states only, emerging from TP$_{1}$ and TP$_{2}$ separately without the bulk states.}
  \label{fig:Ss}
\end{figure}

\subsection{Spin Fluctuations and Multiorbital Susceptibilities}
\begin{figure}
  \includegraphics[width=15 cm]{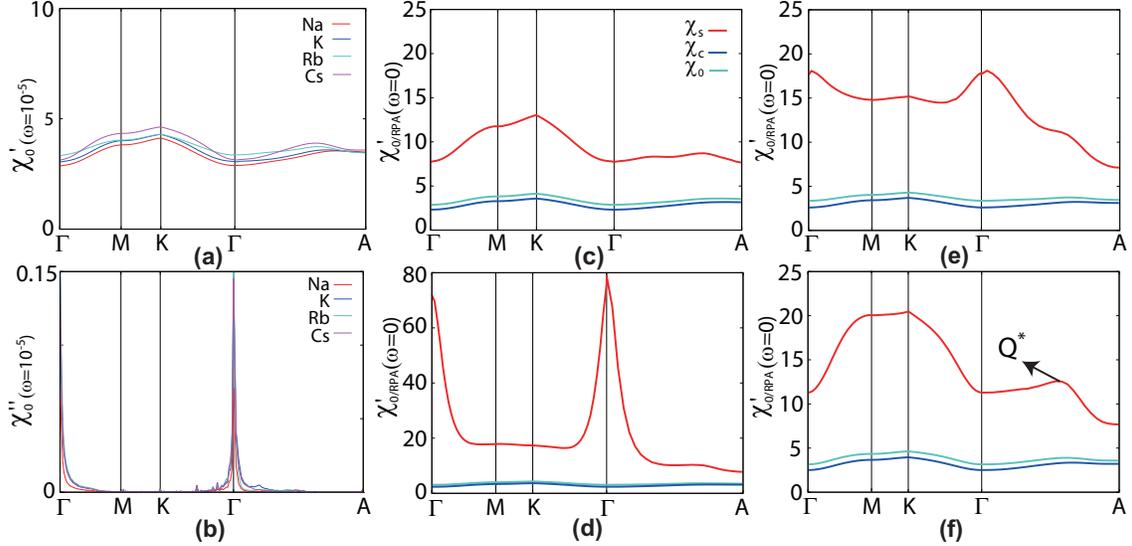}
  \caption{Electron susceptibility of $A_2$Cr$_3$As$_3$. (a) Real and (b) imaginary part of the bare electron susceptibility of the $A_2$Cr$_3$As$_3$ family. (c-f) the real part of magnetic susceptibility calculated at the RPA level for (c) Na$_2$Cr$_3$As$_3$, (d) K$_2$Cr$_3$As$_3$, (e) Rb$_2$Cr$_3$As$_3$ and (f) Cs$_2$Cr$_3$As$_3$, respectively. The red, blue and cyan solid line are spin, charge and bare (in comparison) susceptibilities, respectively, along high symmetry lines for the compounds of this family with $U=2.0$ eV and $J=0.3$ eV.}
  \label{fig:chi_high_sym_line}
\end{figure}

The imaginary part of the bare electron susceptibility $\chi_{0}$ of K$_{2}$Cr$_{3}$As$_{3}$ exhibited a strong peak at the $\Gamma$ point\cite{Jiang2015}. In order to investigate the variation of electron susceptibility, we also calculate susceptibility $\chi_{0}$ using the Lindhard response function for all compounds of this family. As shown in Fig.~\ref{fig:chi_high_sym_line}(a-b), the bare susceptibilities of Na$_{2}$Cr$_{3}$As$_{3}$, Rb$_{2}$Cr$_{3}$As$_{3}$ and Cs$_{2}$Cr$_{3}$As$_{3}$ resemble that of K$_{2}$Cr$_{3}$As$_{3}$. Specifically, the real part $\chi_{0}'$ is relatively featureless, while the imaginary part $\chi_{0}''$ is dominated by a resonance peak at the $\Gamma$ point. As reported in previous NMR studies \cite{NMR_K233,NMR_Rb233,NMR_Cs233}, there were clear differences in the Knight shift and NMR relaxation rate ($1/T_{1}T$) of K$_{2}$Cr$_{3}$As$_{3}$ (Rb$_{2}$Cr$_{3}$As$_{3}$) and Cs$_{2}$Cr$_{3}$As$_{3}$, implying that the spin susceptibility of K$_{2}$Cr$_{3}$As$_{3}$ (Rb$_{2}$Cr$_{3}$As$_{3}$) strikingly differs from that of Cs$_{2}$Cr$_{3}$As$_{3}$. Therefore, electron-electron interactions must be taken into consideration to explain the aforementioned NMR experiments. We calculate both spin and charge susceptibilities within random phase approximation (RPA) with intra-orbital Coulomb ($U$), inter-orbital Coulomb ($U'$), Hund's coupling ($J$) and pair-hopping ($J'$) interactions involved (for details see Supplementary Method). In Fig.~\ref{fig:chi_high_sym_line}(c-f), the real part of charge (spin) susceptibility $\chi_{c}'$ ($\chi_{s}'$ ) is presented for $U=2$ eV and $J=0.3$ eV in comparison to that of bare ones. For all members in this family, the spin susceptibilities are significantly enhanced, while the charge susceptibilities are suppressed. The sharp peak present at the $\Gamma$ point of K$_{2}$Cr$_{3}$As$_{3}$ indicates that it is very close to the critical point with certain values of $U$ and $J$. Naively, using the simplest approximation, the enhancement of the imaginary part $\chi''$ can be written as $\chi''(\mathbf{q},\omega)\approx\chi_{0}(\mathbf{q},\omega)/[1-\overline{U}\chi'_{0}(\mathbf{q},\omega)]^{2}$\cite{Mazin_NMR_MgB2,PhysRevLett.72.1933}, where $\overline{U}$ is the Stoner factor. Applying this approximation, the imaginary part $\chi''$ of K$_{2}$Cr$_{3}$As$_{3}$ is also expected to be strongly enhanced at the $\Gamma$ point when $1-U\chi'_{0}(\mathbf{q},\omega)$ approaches zero. The scattering peak of $\chi'$ in Rb$_{2}$Cr$_{3}$As$_{3}$ still locates at the $\Gamma$ point but is lower and broader compared to K$_{2}$Cr$_{3}$As$_{3}$. In the case of Cs$_{2}$Cr$_{3}$As$_{3}$ (Na$_{2}$Cr$_{3}$As$_{3}$), in contrast, only a broadened and plateau-like structure can be found around the $M(\pi, 0, 0)$ and $K(\frac{2}{3}\pi$,$\frac{2}{3}\pi,0)$ points. Interestingly, there exists another broad hump around $Q^{\ast}=(0,0,0.6\pi)$ in Cs$_{2}$Cr$_{3}$As$_{3}$, which is possibly due to intraband scattering within the $\gamma$ band. Similar to previous report in iron-based superconductor LaOFeAs\cite{LaFeAsO_Eermin}, such a broad hump might be related to spin-density-wave (SDW) fluctuations. In K$_{2}$Cr$_{3}$As$_{3}$ (Rb$_{2}$Cr$_{3}$As$_{3}$), the SDW hump at $Q^{\ast}$ also exists but not apparent, overwhelmed by the large peak located at $\mathbf{q}(0,0,0)$. It is worth noting that the strong peak of $\chi'$ at the $\Gamma$ point of K$_{2}$Cr$_{3}$As$_{3}$ (Rb$_{2}$Cr$_{3}$As$_{3}$) is robust against the Hund's coupling ($J$). It will also be present even if the Hund's coupling is completely turned off ($J=0$). In addition, the spin susceptibility appears to be insensitive to either the inter-orbital Coulomb repulsion ($U'$) or the pair-hopping interaction ($J'$). 

\begin{figure}
  \includegraphics[width=15 cm]{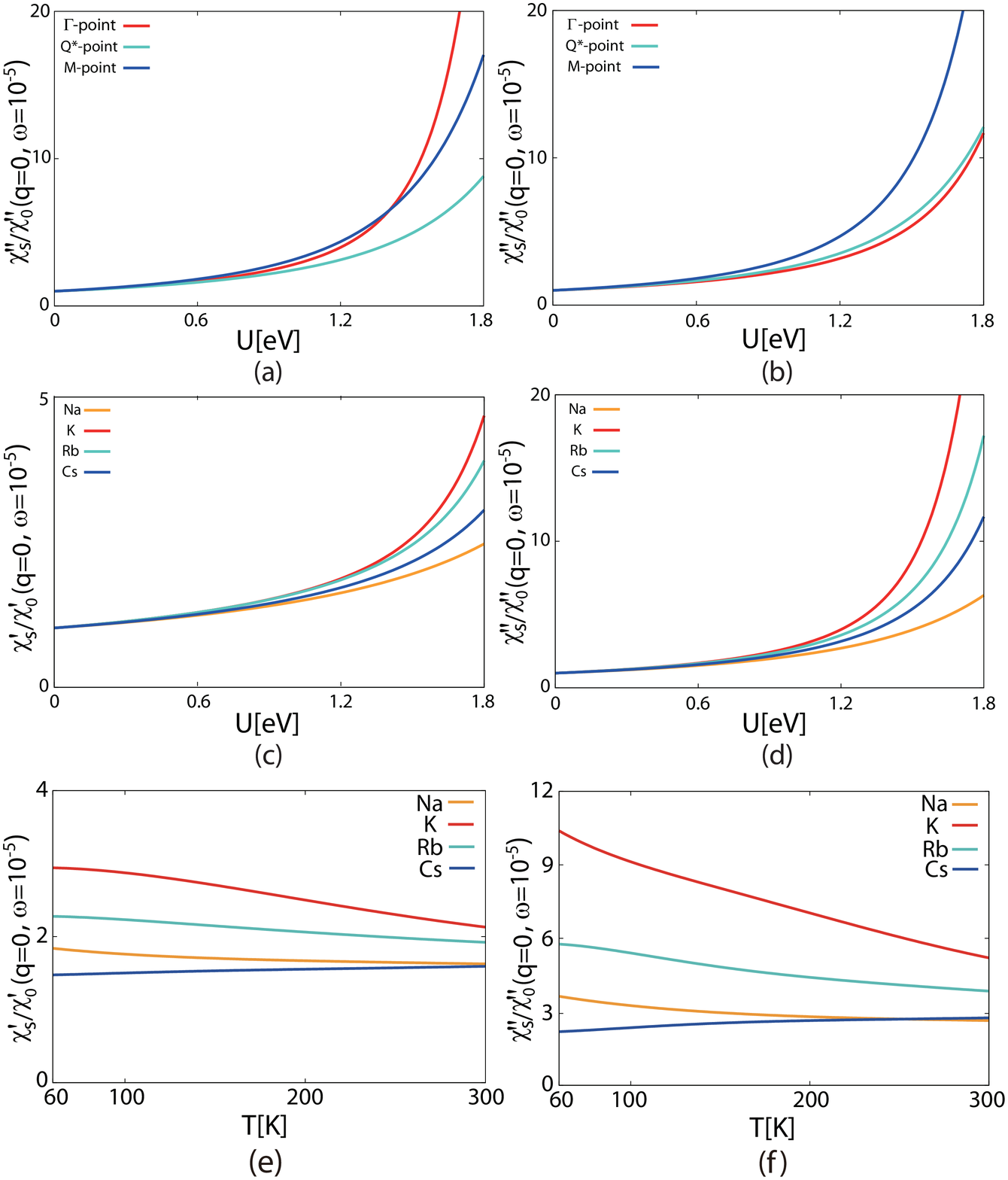}
  \caption{Comparison of spin susceptibility. (a-b) Imaginary part of the spin susceptibility ($\chi_{s}/\chi_{0}$) at $\Gamma$, $M$ and $Q^{\ast}$ (marked in Fig. \ref{fig:chi_high_sym_line} (f)) for (a) K$_2$Cr$_3$As$_3$ and (b) Cs$_2$Cr$_3$As$_3$, respectively. (c-d) show (c)  the real and (d) imaginary part of the spin susceptibility ($\chi_{s}/\chi_{0}$) at $\Gamma$ as a function of intra-orbital Coulomb interaction $U$ for $A_2$Cr$_3$As$_3$ ($A$=Na,K,Rb and Cs). (e-f) Real and imaginary part of the spin susceptibility ($\chi_{s}/\chi_{0}$) as a function of temperature at $\Gamma$ of $A_2$Cr$_3$As$_3$. The interaction parameters have been chosen to be $U=1.2$ eV and $J=0.2U$, and the results for other interaction region ($U<U_{c}$) are similar. }
  \label{fig:U-T-dep}
\end{figure}

Considering the striking difference in $\chi_{s}'$ between K$_{2}$Cr$_{3}$As$_{3}$ and Cs$_{2}$Cr$_{3}$As$_{3}$, we analyze the dynamic spin susceptibility of these two compounds at the $\Gamma$, $M$ and $Q^{\ast}$ point as shown in Fig.~\ref{fig:U-T-dep}(a-b). The spin susceptibility exhibits a larger enhancement at the $\Gamma$ point in K$_2$Cr$_3$As$_3$, while in Cs$_2$Cr$_3$As$_3$ the spin susceptibility seems to be more enhanced at a finite $\mathbf{q}$. For all members in this family, we further compare the enhancement of spin susceptibility $\chi_{s}'$ and $\chi_{s}''$ at $\mathbf{q}$=(0,0,0) in Fig.~\ref{fig:U-T-dep}(c-d). The results show that the enhancement is most significant in K$_2$Cr$_3$As$_3$, followed by Rb$_2$Cr$_3$As$_3$ and Cs$_2$Cr$_3$As$_3$. Such a systematic trend can also be found in the temperature dependent spin susceptibility enhancement (Fig.~\ref{fig:U-T-dep}(e-f)). Note that the enhancement of $\chi''$ is always larger than that of $\chi'$  for both the $T$- and $U$-dependent spin susceptibilities, indicating that the system is away from an FM ordered state.

Intriguingly, if we assume that the NMR relaxation rate $1/T_{1}T$ \footnote{NMR relaxation rate $1/T_{1}T$ is proportional to $\frac{1}{N}\sum\limits_{\mathbf{q}}\frac{A(\mathbf{q})\mathrm{Im}[\chi(\mathbf{q},\omega,T)]}{\omega}$  and $A(\mathbf{q})$ is a geometrical structure factor. Since the  geometrical structure factor has little effect on $1/T_{1}T$ of  LaOFeAs in Ref. \cite{Graser_2009}, we choose $A(\mathbf{q})=1$. The  nuclear magnetic resonance frequency $\omega$ is close to zero,  which is chosen to be $1 \times 10^{-5}$eV in our calculation.} is dominated by the imaginary part of the spin susceptibility at $\mathbf{q}=0$ (or the so-called long wavelength approximation\cite{Moriya_fm,moriya1974nuclear}), the calculated $1/T_{1}T$ for Na$_2$Cr$_3$As$_3$, K$_2$Cr$_3$As$_3$ and Rb$_2$Cr$_3$As$_3$ (Fig.~\ref{fig:U-T-dep}(e-f)) are qualitatively similar to the NMR experiments of \cite{NMR_K233,NMR_Rb233,Zheng_QCP}, although such a Curie-Weiss-like temperature dependence is more accurate within a self-consistent renormalization (SCR) theory \cite{moriya1985physical}. Nevertheless, in contrast to K$_2$Cr$_3$As$_3$ and Rb$_2$Cr$_3$As$_3$, the enhancement of the spin susceptibility in Cs$_2$Cr$_3$As$_3$ (Fig.~\ref{fig:U-T-dep}(e-f)) exhibits a weak temperature dependence. This implies that the large enhancement of the spin susceptibility at the $\Gamma$ point is substantially suppressed in Cs$_2$Cr$_3$As$_3$, consistent with Ref.~\cite{NMR_Cs233}. Similar $T$-dependence of the spin susceptibility were also obtained by Graser \textit{et al.} in LaFeAsO\cite{Graser_2009} and 26\% Co-doped BaFe$_2$As$_2$ \cite{BaFeAsCo_Ning}.

\begin{figure}
 \includegraphics[width=12 cm]{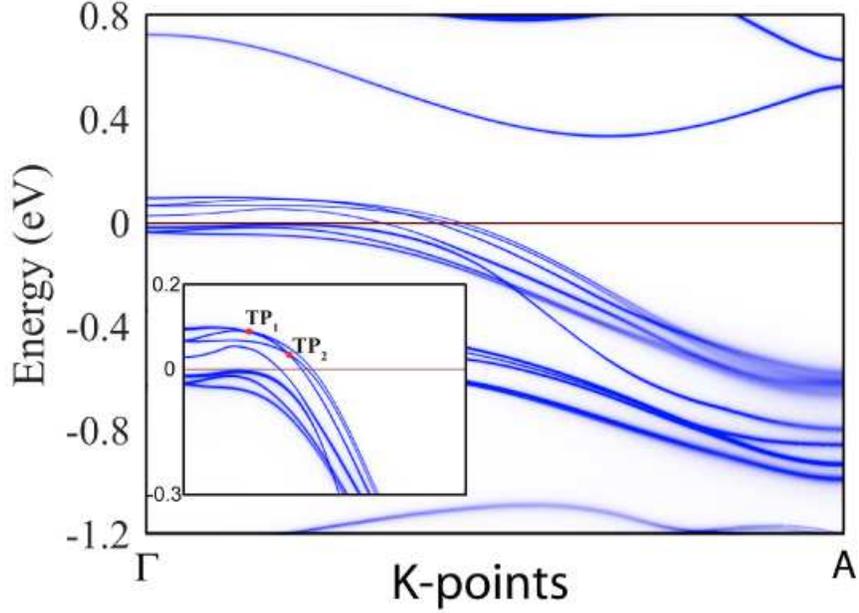}
  \caption{The RPA renormalized quasi-particle spectrum of K$_2$As$_3$Cr$_3$ based on PBE results. Around the Fermi level, the dynamic correlation has relative small effect on the quasi-particle spectrum. The inset shows the same spectrum with smaller energy range. The triply degenerate points are marked as red solid dots. }
  \label{fig:RPA-Bands}
\end{figure}

It is also informative to consider the impact of FM fluctuation on the electronic band structure. As the spin fluctuation is most significant for K$_2$Cr$_3$As$_3$ in our calculations, we calculated its dynamic RPA self-energy $\Sigma(\omega)$ (see Supplementary Figure 4), and obtained its dynamic correlated electronic spectrum. In Fig.~\ref{fig:RPA-Bands}, we show the renormalized quasi-particle energy spectrum. Away from the Fermi level, we see significantly increased scattering rate, most prominent close to A around $E_F-0.4$ eV. However, the renormalization to the quasi-particle states close to the Fermi level is negligible within RPA. Thus, the TPs remain, and are only slightly shifted in K$_{2}$Cr$_{3}$As$_{3}$. Our results are in agreement with previous theoretical studies that correlation effect may be greatly reduced due to the formation of molecular orbitals\cite{ZHOU2017208,Hu_K233,zhong_K233,miao_K233}, and are in line with the experimental observation that no appreciable magnetic local moment is found in these systems. In addition, the comparison between the RPA results and PBE results also illustrates that the dynamic RPA self-energy has little effect on the existence of TPs in these systems. Despite of its moderate effect at normal state, the FM fluctuation would influence the topological properties in superconducting state by allowing for the nontrivial spin-triplet pairing (p- or f-wave), as has been conjectured theoretically \cite{ZHOU2017208,Hu_K233,zhong_K233,miao_K233,zhang2016revisitation} and investigated experimentally\cite{NMR_K233,NMR_Rb233,Zhu_Hc2,pang2015evidence,pang2016penetration}.

\section*{Discussion}
In the normal state, the triply degenerate points topological metal has been regarded as the intermediate phase between Dirac and Weyl semimetals \cite{TP3_typeA}. The TPs can split into Weyl points by breaking $\sigma_{v}$ containing the $C_{3}$ axis or merge into Dirac point after imposing inversion symmetry. Although the 233-type $A_2$Cr$_3$As$_3$ is absent of an inversion center, the 133-type $A$Cr$_3$As$_3$ is centrosymmetric. It is interesting to notice that a Dirac-like crossing point appears between the $\alpha$ and $\beta$ bands along $k_{z}$ in the recently reported KHCr$_3$As$_3$ \cite{taddei2019tuning,Wu_KHCrAs}. In Na$_2$Cr$_3$As$_3$, the surface state lies only 8 meV above the Fermi level, and it is possible for this surface state to become superconducting through proximity effect. In experiment, such a surface superconducting state might have influence on the in-plane $H_{c2}$ \cite{Hc2_UPt3}, although such a possibility has been ruled out in K$_2$Cr$_3$As$_3$ \cite{Zhu_Hc2,cao2018superconductivity}. One possible reason is that the surface state is far away from the Fermi level in stoichiometric K$_2$Cr$_3$As$_3$.

In the superconducting state, lacking inversion symmetry gives rise to Rashba-type ASOC interactions so that the single-particle Hamiltonian takes the form $h(\mathbf{k})=\xi(\mathbf{k})\sigma_{0}+\mathbf{g}_{\mathbf{k}}\cdot\mathbf{\sigma}$. With an extra parity-breaking term $\mathbf{g}_{\mathbf{k}}\cdot\mathbf{\sigma}$, the mixture of singlet and triplet pairing is allowed and the order parameter takes the general form $\Delta(\mathbf{k})=\Psi_{\mathbf{k}}\sigma_{0}+\mathbf{d}_{\mathbf{k}}\cdot\mathbf{\sigma}$, where $\Psi_{\mathbf{k}}=\Psi_{\mathbf{-k}}$ and $\mathbf{d}_{\mathbf{k}}=-\mathbf{d}_{\mathbf{-k}}$. The triplet pairing state can be stable as long as $\mathbf{d}_{\mathbf{k}}$ is parallel to $\mathbf{g}_{\mathbf{k}}$\cite{ASOC_Frigeri}.  {In the following, we discuss the possible nontrivial topological property in superconducting state by the symmetry analysis of superconducting gap $\Delta(\mathbf{k})$ and the spin-orbit coupling $\mathbf{g}_{\mathbf{k}}$. Firstly considering the gap symmetry, the large FM fluctuations favor the spin-triplet pairing component than the spin-singlet one. For the gap symmetry belongs to the $A''$ representation of $D_{3h}$ group (e.g. $p_{z}$-wave \cite{Hu_K233}),  $\Delta(\mathbf{k})$ satisfies $U_{\mathbf{k}}(S_6)\Delta(\mathbf{k})U_{\mathbf{-k}}(S_6)^{T}=\chi_{S_6}\Delta_{S_6}(\mathbf{k})$, where $\chi_{S_6}=-1$ and the six-fold rotoinversion symmetry $S_6=iC_6$  . We calculate the $\mathbb{Z}_6$ indexes ($z_{6}'$ and $z_{6}''$) associated with the $S_6$ according to the symmetry indicators for topological superconductors defined in Refs.\cite{Ono_symmetry_indicators,ono_arxiv_2019refined}. The resulting $z_{6}'$ and $z_{6}''$ are both nonzeros ($z_{6}'=1$ and $z_{6}''=-1$) for mBJ results of Na$_2$Cr$_3$As$_3$ and K$_2$Cr$_3$As$_3$ (for details see Supplementary Table 1). Secondly considering the large 3D FS around the $\Gamma$ point in  K$_2$Cr$_3$As$_3$ (Rb$_2$Cr$_3$As$_3$), the corresponding representation of $D_{3h}$ is $\Gamma_{7}$ ($\Gamma_{8}$), yielding $\mathbf{g}_{\mathbf{k}}=\beta_{1}k_{z}[(k_{x}^{2}-k_{y}^{2})\sigma_{x}-2k_{x}k_{y}\sigma_{y}]+\beta_{2}k_{x}(k_{x}^{2}-3k_{y}^{2})\sigma_{z}$ \cite{smidman2017superconductivity}, where $\beta_1$ and $\beta_2$ are linear combination coefficients. Moreover, the ASOC splitting around $K(\frac{2}{3}\pi$,$\frac{2}{3}\pi,0)$ is roughly 60 meV \cite{Jiang2015}, while along $A(0, 0, \pi)$-$H(\pi, 0, \pi)$ it is nearly negligible. Therefore the ASOC splitting should be dominated by the $\beta_2$ term, i.e., $\mathbf{g}_{\mathbf{k}}= k_{x}(k_{x}^{2}-3k_{y}^{2})\sigma_{z}$ and thus it is invariant under $\mathbf{g}_{\mathbf{k}_{i},+k_{0}}=\mathbf{g}_{\mathbf{k}_{i},-k_{0}}$. It has been demonstrated by Schnyder $et$ $al.$ \cite{Schnyder_NSC} that the presence of such type of symmetry gives rise to topological nontrivial features of line nodes in the gap of NCSs. More importantly, the Majorana surface states can exist at time-reversal-invariant momenta of the surface Brillouin zone. Since in the non-superconducting state the 1D arc surface states connecting the TPs already exist, it will be more interesting to further investigate the topological properties in the superconducting state of $A_2$Cr$_3$As$_3$.

In summary, we have performed first-principles calculations on $A_2$Cr$_3$As$_3$ ($A$=Na, K, Rb and Cs) and analyzed the systematic variations of electronic structures, FSs, topological properties and magnetic spin susceptibilities of the compounds in this family. Using the surface Green's function method we calculate the (010) surface state and find the TPs in Na$_2$Cr$_3$As$_3$ and K$_2$Cr$_3$As$_3$ are connected by double 1D Fermi arcs. To explore the behavior of magnetic spin response, charge and spin susceptibility are obtained by employing RPA calculations. We demonstrate that the strong enhancement of spin fluctuations is present at the $\Gamma$ point in K$_2$Cr$_3$As$_3$ and Rb$_2$Cr$_3$As$_3$, while in Cs$_2$Cr$_3$As$_3$ and Na$_2$Cr$_3$As$_3$, the FM spin fluctuations are not apparently enhanced. The existence of TPs is robust against the FM spin fluctuation under RPA. Based on symmetry analysis, the $A_2$Cr$_3$As$_3$ compounds are possible candidates for topological superconductor.

\emph{Note added}. During the preparation of our manuscript, we became aware of a recent paper\cite{Liu_233}. In addition to the scenario we discussed here, they have proposed that triplet $p_{z}$-pairing can also be topological superconductors with weak SOC.

\section{Methods}
 \subsection{Calculation Parameters}
The calculations were carried out using density functional theory (DFT) as implemented in the Vienna Abinitio Simulation Package (VASP)\cite{Kresse_1993,Kresse_1999}. The Perdew, Burke and Ernzerhoff parameterization (PBE) of generalized gradient approximation (GGA) to the exhange correlation functional was employed\cite{PBE}. The energy cutoff of the plane-wave basis was up to 450 eV, and 6$\times$6$\times$12 $\Gamma$-centered Monkhorst-Pack \cite{Monkhorst-Pack} k-point mesh was chosen to ensure that the total energy converges to 1 meV/cell. We also performed a comparison with the modified Becke–Johnson exchange potentials (mBJ)\cite{mBJ} for Na$_{2}$Cr$_{3}$As$_{3}$ and K$_{2}$Cr$_{3}$As$_{3}$.  

 \subsection{Structural Parameters}
The calculations of K$_{2}$Cr$_{3}$As$_{3}$, Rb$_{2}$Cr$_{3}$As$_{3}$ and Cs$_{2}$Cr$_{3}$As$_{3}$ were performed with experimental crystal structure. For Na$_{2}$Cr$_{3}$As$_{3}$, only the lattice constants are available from experiment, therefore we employed its lattice constants from experiment, while the atomic positions are fully relaxed with dynamical mean-field theory (DMFT) at 300K \cite{DMFT_1996,DMFT_2006}.
 
 \subsection{Tight Binding Hamiltonian}
The band structures obtained with the PBE and the mBJ methods were fitted to a TB model Hamiltonian with maximally projected Wannier function method\cite{wannier90,MPWF} using 48 atomic orbitals including Cr 3$d$ and As 4$p$. The resulting Hamiltonian was then used to calculate the Fermi surface, charge (spin) susceptibility, self-energy, and surface state with surface Green’s function\cite{sancho1985highly}.

\section*{Code availability}
The computer code to perform RPA calculations is available from the corresponding author upon reasonable request.

\section*{Data availability}
The data that support the findings of this study are available from the corresponding author upon reasonable request.

\section*{Acknowledgments}
The authors would like to thank Yi Zhou, Guanghan Cao, Jianhui Dai, Si-Qi Wu, and Xiang Lv for the inspiring discussions. The calculations were partly performed at the Tianhe-2 National Supercomputing Center in China and the HPC center at Hangzhou Normal University. This work has been supported by the NSFC (Nos. 11874137, 11574265, and 11774309) and the 973 project (Nos. 2014CB648400 and 2016YFA0300402). 

\section*{Competing Interests}
The authors declare no competing interests.

\section*{Author contributions}
C.C.X. and C. C. designed research; C.C.X. performed the calculations; C.C.X., N.H.W, Q.C and C. C. drafted the manuscript; C.C.X. and C. C. were responsible for the data analysis; All the authors participated in discussions.


\begin{thebibliography}{10}
\expandafter\ifx\csname url\endcsname\relax
  \def\url#1{\texttt{#1}}\fi
\expandafter\ifx\csname urlprefix\endcsname\relax\def\urlprefix{URL }\fi
\providecommand{\bibinfo}[2]{#2}
\providecommand{\eprint}[2][]{\url{#2}}

\bibitem{Qi_RMP}
\bibinfo{author}{Qi, X.-L.} \& \bibinfo{author}{Zhang, S.-C.}
\newblock \bibinfo{title}{Topological insulators and superconductors}.
\newblock \emph{\bibinfo{journal}{Rev. Mod. Phys.}}
  \textbf{\bibinfo{volume}{83}}, \bibinfo{pages}{1057--1110}
  (\bibinfo{year}{2011}).

\bibitem{Bi2Se3_zhang}
\bibinfo{author}{Zhang, H.} \emph{et~al.}
\newblock \bibinfo{title}{Topological insulators in Bi$_2$Se$_3$, Bi$_2$Te$_3$ and Sb$_2$Te$_3$ with a single dirac cone on the surface}.
\newblock \emph{\bibinfo{journal}{Nature physics}}
  \textbf{\bibinfo{volume}{5}}, \bibinfo{pages}{438} (\bibinfo{year}{2009}).

\bibitem{Fu_RMP}
\bibinfo{author}{Hasan, M.~Z.} \& \bibinfo{author}{Kane, C.~L.}
\newblock \bibinfo{title}{Colloquium: Topological insulators}.
\newblock \emph{\bibinfo{journal}{Rev. Mod. Phys.}}
  \textbf{\bibinfo{volume}{82}}, \bibinfo{pages}{3045--3067}
  (\bibinfo{year}{2010}).

\bibitem{Wan_Weyl}
\bibinfo{author}{Wan, X.}, \bibinfo{author}{Turner, A.~M.},
  \bibinfo{author}{Vishwanath, A.} \& \bibinfo{author}{Savrasov, S.~Y.}
\newblock \bibinfo{title}{Topological semimetal and fermi-arc surface states in
  the electronic structure of pyrochlore iridates}.
\newblock \emph{\bibinfo{journal}{Phys. Rev. B}} \textbf{\bibinfo{volume}{83}},
  \bibinfo{pages}{205101} (\bibinfo{year}{2011}).

\bibitem{Weng_Weyl}
\bibinfo{author}{Weng, H.}, \bibinfo{author}{Fang, C.}, \bibinfo{author}{Fang,
  Z.}, \bibinfo{author}{Bernevig, B.~A.} \& \bibinfo{author}{Dai, X.}
\newblock \bibinfo{title}{Weyl semimetal phase in noncentrosymmetric
  transition-metal monophosphides}.
\newblock \emph{\bibinfo{journal}{Phys. Rev. X}} \textbf{\bibinfo{volume}{5}},
  \bibinfo{pages}{011029} (\bibinfo{year}{2015}).

\bibitem{Lv_TaAs}
\bibinfo{author}{Lv, B.~Q.} \emph{et~al.}
\newblock \bibinfo{title}{Experimental discovery of weyl semimetal TaAs}.
\newblock \emph{\bibinfo{journal}{Phys. Rev. X}} \textbf{\bibinfo{volume}{5}},
  \bibinfo{pages}{031013} (\bibinfo{year}{2015}).

\bibitem{Na3Bi_XiDaiZhongFan}
\bibinfo{author}{Wang, Z.} \emph{et~al.}
\newblock \bibinfo{title}{Dirac semimetal and topological phase transitions in
  ${A}_{3}$Bi ($A=\text{Na}$, K, Rb)}.
\newblock \emph{\bibinfo{journal}{Phys. Rev. B}} \textbf{\bibinfo{volume}{85}},
  \bibinfo{pages}{195320} (\bibinfo{year}{2012}).

\bibitem{Cd3As2_XiDaiZhongFang}
\bibinfo{author}{Wang, Z.}, \bibinfo{author}{Weng, H.}, \bibinfo{author}{Wu,
  Q.}, \bibinfo{author}{Dai, X.} \& \bibinfo{author}{Fang, Z.}
\newblock \bibinfo{title}{Three-dimensional dirac semimetal and quantum
  transport in Cd${}_{3}$As${}_{2}$}.
\newblock \emph{\bibinfo{journal}{Phys. Rev. B}} \textbf{\bibinfo{volume}{88}},
  \bibinfo{pages}{125427} (\bibinfo{year}{2013}).

\bibitem{Na3Bi_liu}
\bibinfo{author}{Liu, Z.} \emph{et~al.}
\newblock \bibinfo{title}{Discovery of a three-dimensional topological dirac
  semimetal, Na$_3$Bi}.
\newblock \emph{\bibinfo{journal}{Science}} \textbf{\bibinfo{volume}{343}},
  \bibinfo{pages}{864--867} (\bibinfo{year}{2014}).

\bibitem{Cd3As2_Cava}
\bibinfo{author}{Borisenko, S.} \emph{et~al.}
\newblock \bibinfo{title}{Experimental realization of a three-dimensional dirac
  semimetal}.
\newblock \emph{\bibinfo{journal}{Phys. Rev. Lett.}}
  \textbf{\bibinfo{volume}{113}}, \bibinfo{pages}{027603}
  (\bibinfo{year}{2014}).

\bibitem{Burkov_nodal_line}
\bibinfo{author}{Burkov, A.~A.}, \bibinfo{author}{Hook, M.~D.} \&
  \bibinfo{author}{Balents, L.}
\newblock \bibinfo{title}{Topological nodal semimetals}.
\newblock \emph{\bibinfo{journal}{Phys. Rev. B}} \textbf{\bibinfo{volume}{84}},
  \bibinfo{pages}{235126} (\bibinfo{year}{2011}).

\bibitem{Weng_Cu3PdN}
\bibinfo{author}{Yu, R.}, \bibinfo{author}{Weng, H.}, \bibinfo{author}{Fang,
  Z.}, \bibinfo{author}{Dai, X.} \& \bibinfo{author}{Hu, X.}
\newblock \bibinfo{title}{Topological node-line semimetal and dirac semimetal
  state in antiperovskite ${\mathrm{Cu}}_{3}\mathrm{PdN}$}.
\newblock \emph{\bibinfo{journal}{Phys. Rev. Lett.}}
  \textbf{\bibinfo{volume}{115}}, \bibinfo{pages}{036807}
  (\bibinfo{year}{2015}).

\bibitem{Weng_MTC}
\bibinfo{author}{Weng, H.} \emph{et~al.}
\newblock \bibinfo{title}{Topological node-line semimetal in three-dimensional
  graphene networks}.
\newblock \emph{\bibinfo{journal}{Phys. Rev. B}} \textbf{\bibinfo{volume}{92}},
  \bibinfo{pages}{045108} (\bibinfo{year}{2015}).

\bibitem{TP3_typeA}
\bibinfo{author}{Zhu, Z.}, \bibinfo{author}{Winkler, G.~W.},
  \bibinfo{author}{Wu, Q.}, \bibinfo{author}{Li, J.} \&
  \bibinfo{author}{Soluyanov, A.~A.}
\newblock \bibinfo{title}{Triple point topological metals}.
\newblock \emph{\bibinfo{journal}{Phys. Rev. X}} \textbf{\bibinfo{volume}{6}},
  \bibinfo{pages}{031003} (\bibinfo{year}{2016}).

\bibitem{TP3_InAsSb}
\bibinfo{author}{Winkler, G.~W.}, \bibinfo{author}{Wu, Q.},
  \bibinfo{author}{Troyer, M.}, \bibinfo{author}{Krogstrup, P.} \&
  \bibinfo{author}{Soluyanov, A.~A.}
\newblock \bibinfo{title}{Topological phases in
  ${\mathrm{InAs}}_{1\ensuremath{-}x}{\mathrm{Sb}}_{x}$: From novel topological
  semimetal to majorana wire}.
\newblock \emph{\bibinfo{journal}{Phys. Rev. Lett.}}
  \textbf{\bibinfo{volume}{117}}, \bibinfo{pages}{076403}
  (\bibinfo{year}{2016}).

\bibitem{Schnyder_NSC}
\bibinfo{author}{Schnyder, A.~P.}, \bibinfo{author}{Brydon, P. M.~R.} \&
  \bibinfo{author}{Timm, C.}
\newblock \bibinfo{title}{Types of topological surface states in nodal
  noncentrosymmetric superconductors}.
\newblock \emph{\bibinfo{journal}{Phys. Rev. B}} \textbf{\bibinfo{volume}{85}},
  \bibinfo{pages}{024522} (\bibinfo{year}{2012}).

\bibitem{UPt3_Timm}
\bibinfo{author}{Agterberg, D.~F.}, \bibinfo{author}{Brydon, P. M.~R.} \&
  \bibinfo{author}{Timm, C.}
\newblock \bibinfo{title}{Bogoliubov fermi surfaces in superconductors with
  broken time-reversal symmetry}.
\newblock \emph{\bibinfo{journal}{Phys. Rev. Lett.}}
  \textbf{\bibinfo{volume}{118}}, \bibinfo{pages}{127001}
  (\bibinfo{year}{2017}).

\bibitem{schnyder_nodal_topological_review}
\bibinfo{author}{Schnyder, A.~P.} \& \bibinfo{author}{Brydon, P.~M.}
\newblock \bibinfo{title}{Topological surface states in nodal superconductors}.
\newblock \emph{\bibinfo{journal}{Journal of Physics: Condensed Matter}}
  \textbf{\bibinfo{volume}{27}}, \bibinfo{pages}{243201}
  (\bibinfo{year}{2015}).

\bibitem{PhysRevB.84.060504}
\bibinfo{author}{Schnyder, A.~P.} \& \bibinfo{author}{Ryu, S.}
\newblock \bibinfo{title}{Topological phases and surface flat bands in
  superconductors without inversion symmetry}.
\newblock \emph{\bibinfo{journal}{Phys. Rev. B}} \textbf{\bibinfo{volume}{84}},
  \bibinfo{pages}{060504} (\bibinfo{year}{2011}).

\bibitem{PhysRevB.84.020501}
\bibinfo{author}{Brydon, P. M.~R.}, \bibinfo{author}{Schnyder, A.~P.} \&
  \bibinfo{author}{Timm, C.}
\newblock \bibinfo{title}{Topologically protected flat zero-energy surface
  bands in noncentrosymmetric superconductors}.
\newblock \emph{\bibinfo{journal}{Phys. Rev. B}} \textbf{\bibinfo{volume}{84}},
  \bibinfo{pages}{020501} (\bibinfo{year}{2011}).

\bibitem{matsuura2013protected}
\bibinfo{author}{Matsuura, S.}, \bibinfo{author}{Chang, P.-Y.},
  \bibinfo{author}{Schnyder, A.~P.} \& \bibinfo{author}{Ryu, S.}
\newblock \bibinfo{title}{Protected boundary states in gapless topological
  phases}.
\newblock \emph{\bibinfo{journal}{New Journal of Physics}}
  \textbf{\bibinfo{volume}{15}}, \bibinfo{pages}{065001}
  (\bibinfo{year}{2013}).

\bibitem{pang2015evidence}
\bibinfo{author}{Pang, G.} \emph{et~al.}
\newblock \bibinfo{title}{Evidence for nodal superconductivity in
  quasi-one-dimensional K$_2$Cr$_3$As$_3$}.
\newblock \emph{\bibinfo{journal}{Physical Review B}}
  \textbf{\bibinfo{volume}{91}}, \bibinfo{pages}{220502}
  (\bibinfo{year}{2015}).

\bibitem{pang2016penetration}
\bibinfo{author}{Pang, G.} \emph{et~al.}
\newblock \bibinfo{title}{Penetration depth measurements of K$_2$Cr$_3$As$_3$ and
  Rb$_2$Cr$_3$As$_3$}.
\newblock \emph{\bibinfo{journal}{Journal of Magnetism and Magnetic Materials}}
  \textbf{\bibinfo{volume}{400}}, \bibinfo{pages}{84--87}
  (\bibinfo{year}{2016}).

\bibitem{NMR_K233}
\bibinfo{author}{Zhi, H.~Z.}, \bibinfo{author}{Imai, T.},
  \bibinfo{author}{Ning, F.~L.}, \bibinfo{author}{Bao, J.-K.} \&
  \bibinfo{author}{Cao, G.-H.}
\newblock \bibinfo{title}{Nmr investigation of the quasi-one-dimensional
  superconductor K$_2$Cr$_3$As$_3$}.
\newblock \emph{\bibinfo{journal}{Phys. Rev. Lett.}}
  \textbf{\bibinfo{volume}{114}}, \bibinfo{pages}{147004}
  (\bibinfo{year}{2015}).

\bibitem{NMR_Rb233}
\bibinfo{author}{Yang, J.}, \bibinfo{author}{Tang, Z.~T.},
  \bibinfo{author}{Cao, G.~H.} \& \bibinfo{author}{Zheng, G.-q.}
\newblock \bibinfo{title}{Ferromagnetic spin fluctuation and unconventional
  superconductivity in Rb$_2$Cr$_3$As$_3$
  revealed by $^{75}\mathrm{As}$ NMR and NQR}.
\newblock \emph{\bibinfo{journal}{Phys. Rev. Lett.}}
  \textbf{\bibinfo{volume}{115}}, \bibinfo{pages}{147002}
  (\bibinfo{year}{2015}).

\bibitem{muSR_K}
\bibinfo{author}{Adroja, D.~T.} \emph{et~al.}
\newblock \bibinfo{title}{Superconducting ground state of quasi-one-dimensional
  K$_2$Cr$_3$As$_3$ investigated using
  $\ensuremath{\mu}\text{SR}$ measurements}.
\newblock \emph{\bibinfo{journal}{Phys. Rev. B}} \textbf{\bibinfo{volume}{92}},
  \bibinfo{pages}{134505} (\bibinfo{year}{2015}).

\bibitem{muSR_Cs}
\bibinfo{author}{Adroja, D.} \emph{et~al.}
\newblock \bibinfo{title}{Nodal superconducting gap structure in the
  quasi-one-dimensional Cs$_2$Cr$_3$As$_3$ investigated using $\mu$SR measurements}.
\newblock \emph{\bibinfo{journal}{Journal of the Physical Society of Japan}}
  \textbf{\bibinfo{volume}{86}}, \bibinfo{pages}{044710}
  (\bibinfo{year}{2017}).

\bibitem{shao_K_specific-heat}
\bibinfo{author}{Shao, Y.~T.} \emph{et~al.}
\newblock \bibinfo{title}{Evidence of line nodes in superconducting gap
  function in K$_2$Cr$_3$As$_3$ from specific-heat measurements}.
\newblock \emph{\bibinfo{journal}{{EPL} (Europhysics Letters)}}
  \textbf{\bibinfo{volume}{123}}, \bibinfo{pages}{57001}
  (\bibinfo{year}{2018}).

\bibitem{Rb_2015}
\bibinfo{author}{Tang, Z.-T.} \emph{et~al.}
\newblock \bibinfo{title}{Unconventional superconductivity in
  quasi-one-dimensional Rb$_2$Cr$_3$As$_3$}.
\newblock \emph{\bibinfo{journal}{Phys. Rev. B}} \textbf{\bibinfo{volume}{91}},
  \bibinfo{pages}{020506} (\bibinfo{year}{2015}).

\bibitem{Jiang2015}
\bibinfo{author}{Jiang, H.}, \bibinfo{author}{Cao, G.} \& \bibinfo{author}{Cao,
  C.}
\newblock \bibinfo{title}{Electronic structure of quasi-one-dimensional
  superconductor K$_2$Cr$_3$As$_3$ from first-principles calculations}.
\newblock \emph{\bibinfo{journal}{Scientific Reports}}
  \textbf{\bibinfo{volume}{5}}, \bibinfo{pages}{16054 EP --}
  (\bibinfo{year}{2015}).

\bibitem{Zheng_QCP}
\bibinfo{author}{Luo, J.} \emph{et~al.}
\newblock \bibinfo{title}{Tuning the distance to a possible ferromagnetic
  quantum critical point in $A_2$Cr$_3$As$_3$}.
\newblock \emph{\bibinfo{journal}{Phys. Rev. Lett.}}
  \textbf{\bibinfo{volume}{123}}, \bibinfo{pages}{047001}
  (\bibinfo{year}{2019}).

\bibitem{NMR_Cs233}
\bibinfo{author}{Zhi, H.} \emph{et~al.}
\newblock \bibinfo{title}{$^{133}\mathrm{Cs}$ and $^{75}\mathrm{As}$ nmr
  investigation of the normal metallic state of quasi-one-dimensional
  Cs$_2$Cr$_3$As$_3$}.
\newblock \emph{\bibinfo{journal}{Phys. Rev. B}} \textbf{\bibinfo{volume}{93}},
  \bibinfo{pages}{174508} (\bibinfo{year}{2016}).

\bibitem{Na_2019}
\bibinfo{author}{Mu, Q.-G.} \emph{et~al.}
\newblock \bibinfo{title}{Ion-exchange synthesis and superconductivity at 8.6 K
  of
  $\mathrm{N}{\mathrm{a}}_{2}\mathrm{C}{\mathrm{r}}_{3}\mathrm{A}{\mathrm{s}}_{3}$
  with quasi-one-dimensional crystal structure}.
\newblock \emph{\bibinfo{journal}{Phys. Rev. Materials}}
  \textbf{\bibinfo{volume}{2}}, \bibinfo{pages}{034803} (\bibinfo{year}{2018}).

\bibitem{K_2015}
\bibinfo{author}{Bao, J.-K.} \emph{et~al.}
\newblock \bibinfo{title}{Superconductivity in quasi-one-dimensional
  K$_2$Cr$_3$As$_3$ with significant
  electron correlations}.
\newblock \emph{\bibinfo{journal}{Phys. Rev. X}} \textbf{\bibinfo{volume}{5}},
  \bibinfo{pages}{011013} (\bibinfo{year}{2015}).

\bibitem{wang_K_Rb233}
\bibinfo{author}{Wang, X.} \emph{et~al.}
\newblock \bibinfo{title}{Tunable electronic anisotropy in single-crystal $A_2$Cr$_3$As$_3$($A$= K, Rb) quasi-one-dimensional superconductors}.
\newblock \emph{\bibinfo{journal}{Physical Review B}}
  \textbf{\bibinfo{volume}{92}}, \bibinfo{pages}{020508}
  (\bibinfo{year}{2015}).

\bibitem{Cs_2015}
\bibinfo{author}{Tang, Z.-T.} \emph{et~al.}
\newblock \bibinfo{title}{Superconductivity in quasi-one-dimensional Cs$_2$Cr$_3$As$_3$
  with large interchain distance}.
\newblock \emph{\bibinfo{journal}{Science China Materials}}
  \textbf{\bibinfo{volume}{58}}, \bibinfo{pages}{16--20}
  (\bibinfo{year}{2015}).

\bibitem{Bao_K133}
\bibinfo{author}{Bao, J.-K.} \emph{et~al.}
\newblock \bibinfo{title}{Cluster spin-glass ground state in
  quasi-one-dimensional KCr$_3$As$_3$}.
\newblock \emph{\bibinfo{journal}{Phys. Rev. B}} \textbf{\bibinfo{volume}{91}},
  \bibinfo{pages}{180404} (\bibinfo{year}{2015}).

\bibitem{ren_K133}
\bibinfo{author}{Mu, Q.-G.} \emph{et~al.}
\newblock \bibinfo{title}{Superconductivity at 5K in quasi-one-dimensional
  Cr-based KCr$_3$As$_3$ single crystals}.
\newblock \emph{\bibinfo{journal}{Phys. Rev. B}} \textbf{\bibinfo{volume}{96}},
  \bibinfo{pages}{140504} (\bibinfo{year}{2017}).

\bibitem{tang_Rb_Cs133}
\bibinfo{author}{Tang, Z.-T.} \emph{et~al.}
\newblock \bibinfo{title}{Synthesis, crystal structure and physical properties
  of quasi-one-dimensional $A$Cr$_3$As$_3$ ($A$= Rb, Cs)}.
\newblock \emph{\bibinfo{journal}{Science China Materials}}
  \textbf{\bibinfo{volume}{58}}, \bibinfo{pages}{543--549}
  (\bibinfo{year}{2015}).

\bibitem{ren_Rb133}
\bibinfo{author}{Liu, T.} \emph{et~al.}
\newblock \bibinfo{title}{Superconductivity at 7.3K in the 133-type Cr-based
  RbCr$_3$As$_3$ single crystals}.
\newblock \emph{\bibinfo{journal}{EPL (Europhysics Letters)}}
  \textbf{\bibinfo{volume}{120}}, \bibinfo{pages}{27006}
  (\bibinfo{year}{2018}).

\bibitem{xian2015magnetism}
\bibinfo{author}{Xian-Xin, W.}, \bibinfo{author}{Cong-Cong, L.},
  \bibinfo{author}{Jing, Y.}, \bibinfo{author}{Heng, F.} \&
  \bibinfo{author}{Jiang-Ping, H.}
\newblock \bibinfo{title}{Magnetism in quasi-one-dimensional $A_2$Cr$_3$As$_3$ ($A$= K, Rb) superconductors}.
\newblock \emph{\bibinfo{journal}{Chinese Physics Letters}}
  \textbf{\bibinfo{volume}{32}}, \bibinfo{pages}{057401}
  (\bibinfo{year}{2015}).

\bibitem{HgTe}
\bibinfo{author}{Zaheer, S.} \emph{et~al.}
\newblock \bibinfo{title}{Spin texture on the fermi surface of tensile-strained
  HgTe}.
\newblock \emph{\bibinfo{journal}{Phys. Rev. B}} \textbf{\bibinfo{volume}{87}},
  \bibinfo{pages}{045202} (\bibinfo{year}{2013}).

\bibitem{mBJ}
\bibinfo{author}{Tran, F.} \& \bibinfo{author}{Blaha, P.}
\newblock \bibinfo{title}{Accurate band gaps of semiconductors and insulators
  with a semilocal exchange-correlation potential}.
\newblock \emph{\bibinfo{journal}{Phys. Rev. Lett.}}
  \textbf{\bibinfo{volume}{102}}, \bibinfo{pages}{226401}
  (\bibinfo{year}{2009}).

\bibitem{Mazin_NMR_MgB2}
\bibinfo{author}{Pavarini, E.} \& \bibinfo{author}{Mazin, I.~I.}
\newblock \bibinfo{title}{NMR relaxation rates and knight shifts in
  MgB$_2$}.
\newblock \emph{\bibinfo{journal}{Phys. Rev. B}} \textbf{\bibinfo{volume}{64}},
  \bibinfo{pages}{140504} (\bibinfo{year}{2001}).

\bibitem{PhysRevLett.72.1933}
\bibinfo{author}{Shastry, B.~S.} \& \bibinfo{author}{Abrahams, E.}
\newblock \bibinfo{title}{What does the Korringa ratio measure?}
\newblock \emph{\bibinfo{journal}{Phys. Rev. Lett.}}
  \textbf{\bibinfo{volume}{72}}, \bibinfo{pages}{1933--1936}
  (\bibinfo{year}{1994}).

\bibitem{LaFeAsO_Eermin}
\bibinfo{author}{Korshunov, M.~M.} \& \bibinfo{author}{Eremin, I.}
\newblock \bibinfo{title}{Theory of magnetic excitations in iron-based layered
  superconductors}.
\newblock \emph{\bibinfo{journal}{Phys. Rev. B}} \textbf{\bibinfo{volume}{78}},
  \bibinfo{pages}{140509} (\bibinfo{year}{2008}).

\bibitem{Moriya_fm}
\bibinfo{author}{Moriya, T.} \& \bibinfo{author}{Kawabata, A.}
\newblock \bibinfo{title}{Effect of spin fluctuations on itinerant electron
  ferromagnetism}.
\newblock \emph{\bibinfo{journal}{Journal of the Physical Society of Japan}}
  \textbf{\bibinfo{volume}{34}}, \bibinfo{pages}{639--651}
  (\bibinfo{year}{1973}).

\bibitem{moriya1974nuclear}
\bibinfo{author}{Moriya, T.} \& \bibinfo{author}{Ueda, K.}
\newblock \bibinfo{title}{Nuclear magnetic relaxation in weakly ferro-and
  antiferromagnetic metals}.
\newblock \emph{\bibinfo{journal}{Solid State Communications}}
  \textbf{\bibinfo{volume}{15}}, \bibinfo{pages}{169--172}
  (\bibinfo{year}{1974}).

\bibitem{moriya1985physical}
\bibinfo{author}{Moriya, T.}
\newblock \bibinfo{title}{Physical properties of weakly and nearly ferro-and
  antiferromagnetic metals}.
\newblock In \emph{\bibinfo{booktitle}{Spin Fluctuations in Itinerant Electron
  Magnetism}}, \bibinfo{pages}{82--108} (\bibinfo{publisher}{Springer},
  \bibinfo{year}{1985}).

\bibitem{Graser_2009}
\bibinfo{author}{Graser, S.}, \bibinfo{author}{Maier, T.~A.},
  \bibinfo{author}{Hirschfeld, P.~J.} \& \bibinfo{author}{Scalapino, D.~J.}
\newblock \bibinfo{title}{Near-degeneracy of several pairing channels in
  multiorbital models for the Fe pnictides}.
\newblock \emph{\bibinfo{journal}{New Journal of Physics}}
  \textbf{\bibinfo{volume}{11}}, \bibinfo{pages}{025016}
  (\bibinfo{year}{2009}).

\bibitem{BaFeAsCo_Ning}
\bibinfo{author}{Ning, F.~L.} \emph{et~al.}
\newblock \bibinfo{title}{Contrasting spin dynamics between underdoped and
  overdoped
  $\mathrm{Ba}({\mathrm{Fe}}_{1\ensuremath{-}x}{\mathrm{Co}}_{x}{)}_{2}{\mathrm{As}}_{2}$}.
\newblock \emph{\bibinfo{journal}{Phys. Rev. Lett.}}
  \textbf{\bibinfo{volume}{104}}, \bibinfo{pages}{037001}
  (\bibinfo{year}{2010}).

\bibitem{ZHOU2017208}
\bibinfo{author}{Zhou, Y.}, \bibinfo{author}{Cao, C.} \&
  \bibinfo{author}{Zhang, F.-C.}
\newblock \bibinfo{title}{Theory for superconductivity in alkali chromium
  arsenides $A_2$Cr$_3$As$_3$ ($A$=K, Rb, Cs)}.
\newblock \emph{\bibinfo{journal}{Science Bulletin}}
  \textbf{\bibinfo{volume}{62}}, \bibinfo{pages}{208 -- 211}
  (\bibinfo{year}{2017}).

\bibitem{Hu_K233}
\bibinfo{author}{Wu, X.}, \bibinfo{author}{Yang, F.}, \bibinfo{author}{Le, C.},
  \bibinfo{author}{Fan, H.} \& \bibinfo{author}{Hu, J.}
\newblock \bibinfo{title}{Triplet ${p}_{z}$-wave pairing in
  quasi-one-dimensional $A_2$Cr$_3$As$_3$
  superconductors ($A$=K, Rb, Cs)}.
\newblock \emph{\bibinfo{journal}{Phys. Rev. B}} \textbf{\bibinfo{volume}{92}},
  \bibinfo{pages}{104511} (\bibinfo{year}{2015}).

\bibitem{zhong_K233}
\bibinfo{author}{Zhong, H.}, \bibinfo{author}{Feng, X.-Y.},
  \bibinfo{author}{Chen, H.} \& \bibinfo{author}{Dai, J.}
\newblock \bibinfo{title}{Formation of molecular-orbital bands in a twisted
  hubbard tube: Implications for unconventional superconductivity in
  K$_2$Cr$_3$As$_3$}.
\newblock \emph{\bibinfo{journal}{Phys. Rev. Lett.}}
  \textbf{\bibinfo{volume}{115}}, \bibinfo{pages}{227001}
  (\bibinfo{year}{2015}).

\bibitem{miao_K233}
\bibinfo{author}{Miao, J.-J.}, \bibinfo{author}{Zhang, F.-C.} \&
  \bibinfo{author}{Zhou, Y.}
\newblock \bibinfo{title}{Instability of three-band tomonaga-luttinger liquid:
  Renormalization group analysis and possible application to
  K$_2$Cr$_3$As$_3$}.
\newblock \emph{\bibinfo{journal}{Phys. Rev. B}} \textbf{\bibinfo{volume}{94}},
  \bibinfo{pages}{205129} (\bibinfo{year}{2016}).

\bibitem{zhang2016revisitation}
\bibinfo{author}{Zhang, L.-D.}, \bibinfo{author}{Wu, X.}, \bibinfo{author}{Fan,
  H.}, \bibinfo{author}{Yang, F.} \& \bibinfo{author}{Hu, J.}
\newblock \bibinfo{title}{Revisitation of superconductivity in K$_2$Cr$_3$As$_3$ based
  on the six-band model}.
\newblock \emph{\bibinfo{journal}{EPL (Europhysics Letters)}}
  \textbf{\bibinfo{volume}{113}}, \bibinfo{pages}{37003}
  (\bibinfo{year}{2016}).

\bibitem{Zhu_Hc2}
\bibinfo{author}{Zuo, H.} \emph{et~al.}
\newblock \bibinfo{title}{Temperature and angular dependence of the upper
  critical field in K$_2$Cr$_3$As$_3$}.
\newblock \emph{\bibinfo{journal}{Phys. Rev. B}} \textbf{\bibinfo{volume}{95}},
  \bibinfo{pages}{014502} (\bibinfo{year}{2017}).

\bibitem{taddei2019tuning}
\bibinfo{author}{Taddei, K.~M.} \emph{et~al.}
\newblock \bibinfo{title}{Tuning from frustrated magnetism to superconductivity
  in quasi-one-dimensional KCr$_3$As$_3$ through
  hydrogen doping}.
\newblock \emph{\bibinfo{journal}{Phys. Rev. B}}
  \textbf{\bibinfo{volume}{100}}, \bibinfo{pages}{220503}
  (\bibinfo{year}{2019}).

\bibitem{Wu_KHCrAs}
\bibinfo{author}{Wu, S.-Q.}, \bibinfo{author}{Cao, C.} \& \bibinfo{author}{Cao,
  G.-H.}
\newblock \bibinfo{title}{Lifshitz transition and nontrivial H-doping effect in
  the Cr-based superconductor
  KCr$_3$As$_3$H$_x$}.
\newblock \emph{\bibinfo{journal}{Phys. Rev. B}}
  \textbf{\bibinfo{volume}{100}}, \bibinfo{pages}{155108}
  (\bibinfo{year}{2019}).

\bibitem{Hc2_UPt3}
\bibinfo{author}{Keller, N.}, \bibinfo{author}{Tholence, J.~L.},
  \bibinfo{author}{Huxley, A.} \& \bibinfo{author}{Flouquet, J.}
\newblock \bibinfo{title}{Angular dependence of the upper critical field of the
  heavy fermion superconductor ${\mathrm{UPt}}_{3}$}.
\newblock \emph{\bibinfo{journal}{Phys. Rev. Lett.}}
  \textbf{\bibinfo{volume}{73}}, \bibinfo{pages}{2364--2367}
  (\bibinfo{year}{1994}).

\bibitem{cao2018superconductivity}
\bibinfo{author}{Cao, G.-H.} \& \bibinfo{author}{Zhu, Z.-W.}
\newblock \bibinfo{title}{Superconductivity with peculiar upper critical fields
  in quasi-one-dimensional Cr-based pnictides}.
\newblock \emph{\bibinfo{journal}{Chinese Physics B}}
  \textbf{\bibinfo{volume}{27}}, \bibinfo{pages}{107401}
  (\bibinfo{year}{2018}).

\bibitem{ASOC_Frigeri}
\bibinfo{author}{Frigeri, P.~A.}, \bibinfo{author}{Agterberg, D.~F.},
  \bibinfo{author}{Koga, A.} \& \bibinfo{author}{Sigrist, M.}
\newblock \bibinfo{title}{Superconductivity without inversion symmetry: MnSi
  versus ${\mathrm{C}\mathrm{e}\mathrm{P}\mathrm{t}}_{3}\mathrm{S}\mathrm{i}$}.
\newblock \emph{\bibinfo{journal}{Phys. Rev. Lett.}}
  \textbf{\bibinfo{volume}{92}}, \bibinfo{pages}{097001}
  (\bibinfo{year}{2004}).

\bibitem{Ono_symmetry_indicators}
\bibinfo{author}{Ono, S.}, \bibinfo{author}{Yanase, Y.} \&
  \bibinfo{author}{Watanabe, H.}
\newblock \bibinfo{title}{Symmetry indicators for topological superconductors}.
\newblock \emph{\bibinfo{journal}{Phys. Rev. Research}}
  \textbf{\bibinfo{volume}{1}}, \bibinfo{pages}{013012} (\bibinfo{year}{2019}).

\bibitem{ono_arxiv_2019refined}
\bibinfo{author}{Ono, S.}, \bibinfo{author}{Po, H.~C.} \&
  \bibinfo{author}{Watanabe, H.}
\newblock \bibinfo{title}{Refined symmetry indicators for topological
  superconductors in all space groups}.
\newblock \emph{\bibinfo{journal}{arXiv preprint arXiv:1909.09634}}
  (\bibinfo{year}{2019}).

\bibitem{smidman2017superconductivity}
\bibinfo{author}{Smidman, M.}, \bibinfo{author}{Salamon, M.},
  \bibinfo{author}{Yuan, H.} \& \bibinfo{author}{Agterberg, D.}
\newblock \bibinfo{title}{Superconductivity and spin--orbit coupling in
  non-centrosymmetric materials: A review}.
\newblock \emph{\bibinfo{journal}{Reports on Progress in Physics}}
  \textbf{\bibinfo{volume}{80}}, \bibinfo{pages}{036501}
  (\bibinfo{year}{2017}).

\bibitem{Liu_233}
\bibinfo{author}{{Liu}, C.-C.} \emph{et~al.}
\newblock \bibinfo{title}{{Intrinsic topological superconductivity with exactly
  flat surface bands in the quasi-one-dimensional A$_2$Cr$_3$As$_3$ (A=Na, K,
  Rb, Cs) superconductors}}.
\newblock \emph{\bibinfo{journal}{arXiv e-prints}}
  \bibinfo{pages}{arXiv:1909.00943} (\bibinfo{year}{2019}).

\bibitem{Kresse_1993}
\bibinfo{author}{Kresse, G.} \& \bibinfo{author}{Hafner, J.}
\newblock \bibinfo{title}{Ab initio molecular dynamics for liquid metals}.
\newblock \emph{\bibinfo{journal}{Phys. Rev. B}} \textbf{\bibinfo{volume}{47}},
  \bibinfo{pages}{558--561} (\bibinfo{year}{1993}).

\bibitem{Kresse_1999}
\bibinfo{author}{Kresse, G.} \& \bibinfo{author}{Joubert, D.}
\newblock \bibinfo{title}{From ultrasoft pseudopotentials to the projector
  augmented-wave method}.
\newblock \emph{\bibinfo{journal}{Phys. Rev. B}} \textbf{\bibinfo{volume}{59}},
  \bibinfo{pages}{1758--1775} (\bibinfo{year}{1999}).

\bibitem{PBE}
\bibinfo{author}{Perdew, J.~P.}, \bibinfo{author}{Burke, K.} \&
  \bibinfo{author}{Ernzerhof, M.}
\newblock \bibinfo{title}{Generalized gradient approximation made simple}.
\newblock \emph{\bibinfo{journal}{Phys. Rev. Lett.}}
  \textbf{\bibinfo{volume}{77}}, \bibinfo{pages}{3865--3868}
  (\bibinfo{year}{1996}).

\bibitem{Monkhorst-Pack}
\bibinfo{author}{Monkhorst, H.~J.} \& \bibinfo{author}{Pack, J.~D.}
\newblock \bibinfo{title}{Special points for brillouin-zone integrations}.
\newblock \emph{\bibinfo{journal}{Phys. Rev. B}} \textbf{\bibinfo{volume}{13}},
  \bibinfo{pages}{5188--5192} (\bibinfo{year}{1976}).

\bibitem{DMFT_1996}
\bibinfo{author}{Georges, A.}, \bibinfo{author}{Kotliar, G.},
  \bibinfo{author}{Krauth, W.} \& \bibinfo{author}{Rozenberg, M.~J.}
\newblock \bibinfo{title}{Dynamical mean-field theory of strongly correlated
  fermion systems and the limit of infinite dimensions}.
\newblock \emph{\bibinfo{journal}{Rev. Mod. Phys.}}
  \textbf{\bibinfo{volume}{68}}, \bibinfo{pages}{13--125}
  (\bibinfo{year}{1996}).

\bibitem{DMFT_2006}
\bibinfo{author}{Kotliar, G.} \emph{et~al.}
\newblock \bibinfo{title}{Electronic structure calculations with dynamical
  mean-field theory}.
\newblock \emph{\bibinfo{journal}{Rev. Mod. Phys.}}
  \textbf{\bibinfo{volume}{78}}, \bibinfo{pages}{865--951}
  (\bibinfo{year}{2006}).

\bibitem{wannier90}
\bibinfo{author}{Mostofi, A.~A.} \emph{et~al.}
\newblock \bibinfo{title}{wannier90: A tool for obtaining maximally-localised
  wannier functions}.
\newblock \emph{\bibinfo{journal}{Computer physics communications}}
  \textbf{\bibinfo{volume}{178}}, \bibinfo{pages}{685--699}
  (\bibinfo{year}{2008}).

\bibitem{MPWF}
\bibinfo{author}{Wang, X.}, \bibinfo{author}{Yates, J.~R.},
  \bibinfo{author}{Souza, I.} \& \bibinfo{author}{Vanderbilt, D.}
\newblock \bibinfo{title}{Ab initio calculation of the anomalous hall
  conductivity by wannier interpolation}.
\newblock \emph{\bibinfo{journal}{Phys. Rev. B}} \textbf{\bibinfo{volume}{74}},
  \bibinfo{pages}{195118} (\bibinfo{year}{2006}).

\bibitem{sancho1985highly}
\bibinfo{author}{Sancho, M.~L.}, \bibinfo{author}{Sancho, J.~L.},
  \bibinfo{author}{Sancho, J.~L.} \& \bibinfo{author}{Rubio, J.}
\newblock \bibinfo{title}{Highly convergent schemes for the calculation of bulk
  and surface green functions}.
\newblock \emph{\bibinfo{journal}{Journal of Physics F: Metal Physics}}
  \textbf{\bibinfo{volume}{15}}, \bibinfo{pages}{851} (\bibinfo{year}{1985}).
  
\bibitem{ARPES_K233}
\bibinfo{author}{Watson, M.~D.} \emph{et~al.}
\newblock \bibinfo{title}{Multiband one-dimensional electronic structure and
  spectroscopic signature of tomonaga-luttinger liquid behavior in
  ${\mathrm{K}}_{2}{\mathrm{Cr}}_{3}{\mathrm{As}}_{3}$}.
\newblock \emph{\bibinfo{journal}{Phys. Rev. Lett.}}
  \textbf{\bibinfo{volume}{118}}, \bibinfo{pages}{097002}
  (\bibinfo{year}{2017}).
  
\bibitem{PhysRevB.75.224509}
\bibinfo{author}{Kubo, K.}
\newblock \bibinfo{title}{Pairing symmetry in a two-orbital hubbard model on a
  square lattice}.
\newblock \emph{\bibinfo{journal}{Phys. Rev. B}} \textbf{\bibinfo{volume}{75}},
  \bibinfo{pages}{224509} (\bibinfo{year}{2007}).

\bibitem{PhysRevB.69.104504}
\bibinfo{author}{Takimoto, T.}, \bibinfo{author}{Hotta, T.} \&
  \bibinfo{author}{Ueda, K.}
\newblock \bibinfo{title}{Strong-coupling theory of superconductivity in a
  degenerate hubbard model}.
\newblock \emph{\bibinfo{journal}{Phys. Rev. B}} \textbf{\bibinfo{volume}{69}},
  \bibinfo{pages}{104504} (\bibinfo{year}{2004}).    

\bibitem{Graser_2010}
\bibinfo{author}{Graser, S.} \emph{et~al.}
\newblock \bibinfo{title}{Spin fluctuations and superconductivity in a
  three-dimensional tight-binding model for
  ${\text{BaFe}}_{2}{\text{As}}_{2}$}.
\newblock \emph{\bibinfo{journal}{Phys. Rev. B}} \textbf{\bibinfo{volume}{81}},
  \bibinfo{pages}{214503} (\bibinfo{year}{2010}).

\end{thebibliography}

\newpage

\begin{appendices}
\renewcommand{\thefigure}{{S}\arabic{figure}}
\setcounter{figure}{0}   
\section{Supplementary Results}

\subsection{Fermi surface and Triply degenerate points}
The Fermi surfaces (FSs) of $A_{2}$Cr$_{3}$As$_{3}$ family have been discussed in the main text. Here, we show all the FSs in Fig.~\ref{fig:fs} in more detail. Without SOC, there are three bands ($\alpha$, $\beta$ and $\gamma$) crossing the Fermi level in Na$_2$Cr$_{3}$As$_{3}$, K$_2$Cr$_{3}$As$_{3}$ and Rb$_2$Cr$_{3}$As$_{3}$. In particular, there is one more band around $\Gamma$ point very close to Fermi level (see Fig.~2(b) in the main text), potentially leading to a new FS at the center of Brillouin zone in Rb$_2$Cr$_{3}$As$_{3}$. In Cr$_2$Cr$_{3}$As$_{3}$, the $\gamma$ band is deformed into an Q1D FS and a fourth 3D FS forms around the $\Gamma$ point. With SOC, surprisingly the 3D FS in Cr$_2$Cr$_{3}$As$_{3}$ disappears, only leaving three Q1D FSs.

The electronic structures of $A_{2}$Cr$_{3}$As$_{3}$ along $k_{z}$ are illustrated in Fig.~\ref{fig:bs_all_kz} (a-d). All the compounds of this family host TPs along this high symmetry line. In comparison, Fig.~\ref{fig:bs_all_kz} (e-f) shows the mBJ results of Na$_{2}$Cr$_{3}$As$_{3}$ and K$_{2}$Cr$_{3}$As$_{3}$. As mentioned in the main text, the mBJ band structure of Na$_{2}$Cr$_{3}$As$_{3}$ is similar to that of PBE result and the only difference is the TP$_{1}$ (TP$_{2}$) (Fig. \ref{fig:bs_all_kz} (a)) lies 8 meV above $\epsilon_F$ while 47 meV above $\epsilon_F$ in Fig. \ref{fig:bs_all_kz} (e). In K$_{2}$Cr$_{3}$As$_{3}$, the $\gamma$ band (PBE) is elevated within mBJ calculations. Consequently, two new TPs (TP$_{1}$ and TP$_{2}$) are created while the two original TPs in PBE calculations still exist but lie below $\epsilon_F$ (i.e., TP$_{1}$ in Fig.~\ref{fig:bs_all_kz} (b) is corresponding to TP$_{3}$ in  Fig.~\ref{fig:bs_all_kz} (f)). In PBE calculations, the overwhelming bulk states are around the TPs in K$_{2}$Cr$_{3}$As$_{3}$, Rb$_{2}$Cr$_{3}$As$_{3}$ and Cs$_{2}$Cr$_{3}$As$_{3}$ leading the corresponding surface states submerged into bulk continuum (Fig.~\ref{fig:Ss_PBE} (a-c)). In addition, the TPs in Rb$_{2}$Cr$_{3}$As$_{3}$ stay too close and in Cs$_{2}$Cr$_{3}$As$_{3}$ the TPs lie deep below $\epsilon_F$ in PBE calculations and thus we do not perform mBJ calculations. In Fig.~\ref{fig:Ss_PBE} (d), we show (010) iso-energy surface state plot at $\epsilon_F+0.15$ eV with disappearance of 3D FS within mBJ calculations, which can match well as the previous ARPES measurements\cite{ARPES_K233}. Besides, the iso-energy surface state at $\epsilon_F-0.07$ eV (not shown) is similar, only containing two 1D FSs. As is known, the K$_{2}$Cr$_{3}$As$_{3}$ is very air sensitive, and thus a slight off-stoichiometry may lead the 3D FS missing.
  
\begin{figure}[ht]
  \includegraphics[width=15 cm]{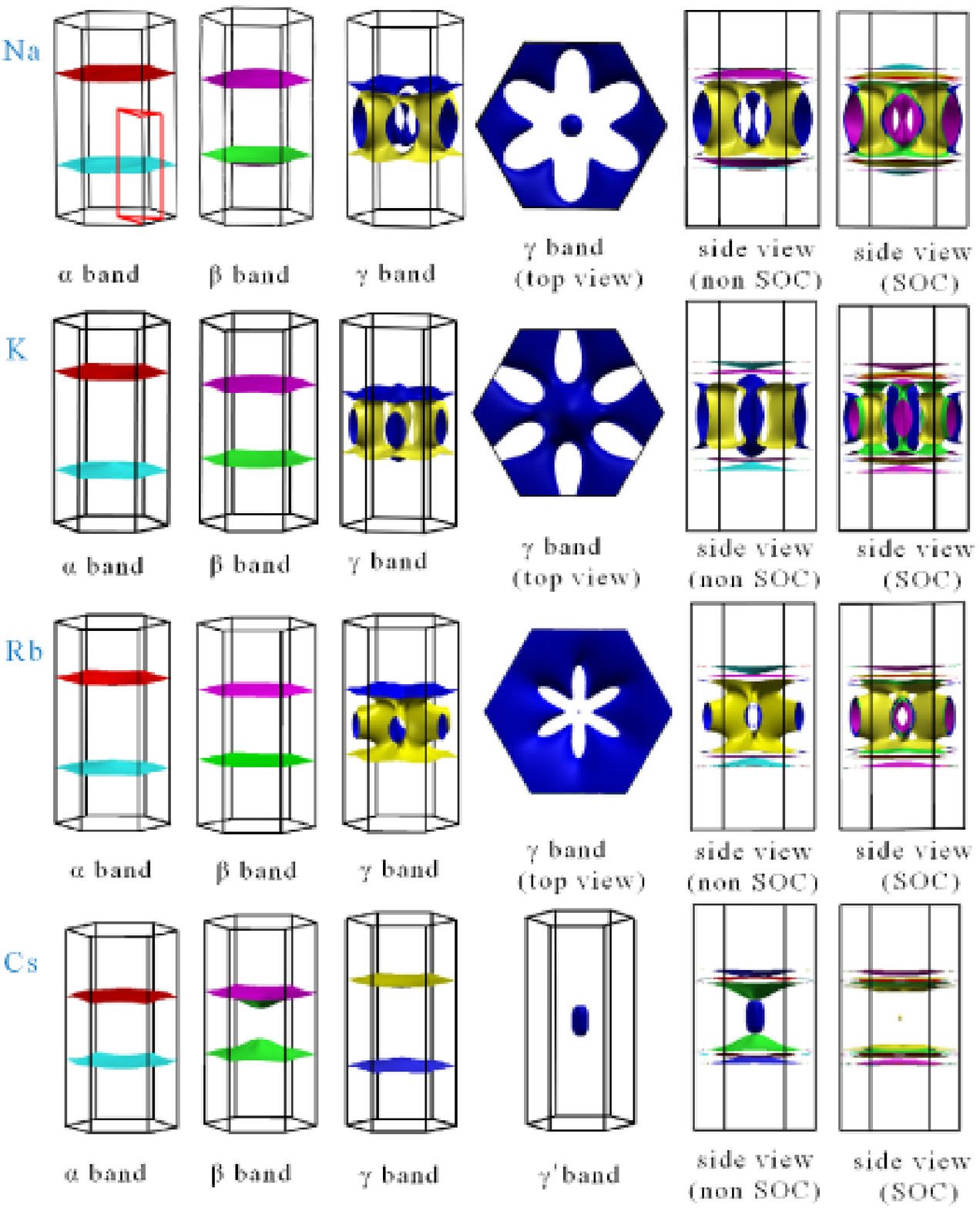}
  \caption{Fermi surface of $A_2$Cr$_3$As$_3$. The panels from top to bottom are
  Na$_2$Cr$_3$As$_3$, K$_2$Cr$_3$As$_3$, Rb$_2$Cr$_3$As$_3$ and Cs$_2$Cr$_3$As$_3$, respectively.
  \label{fig:fs}
  }
\end{figure}

\begin{figure}
 \includegraphics[width=12 cm]{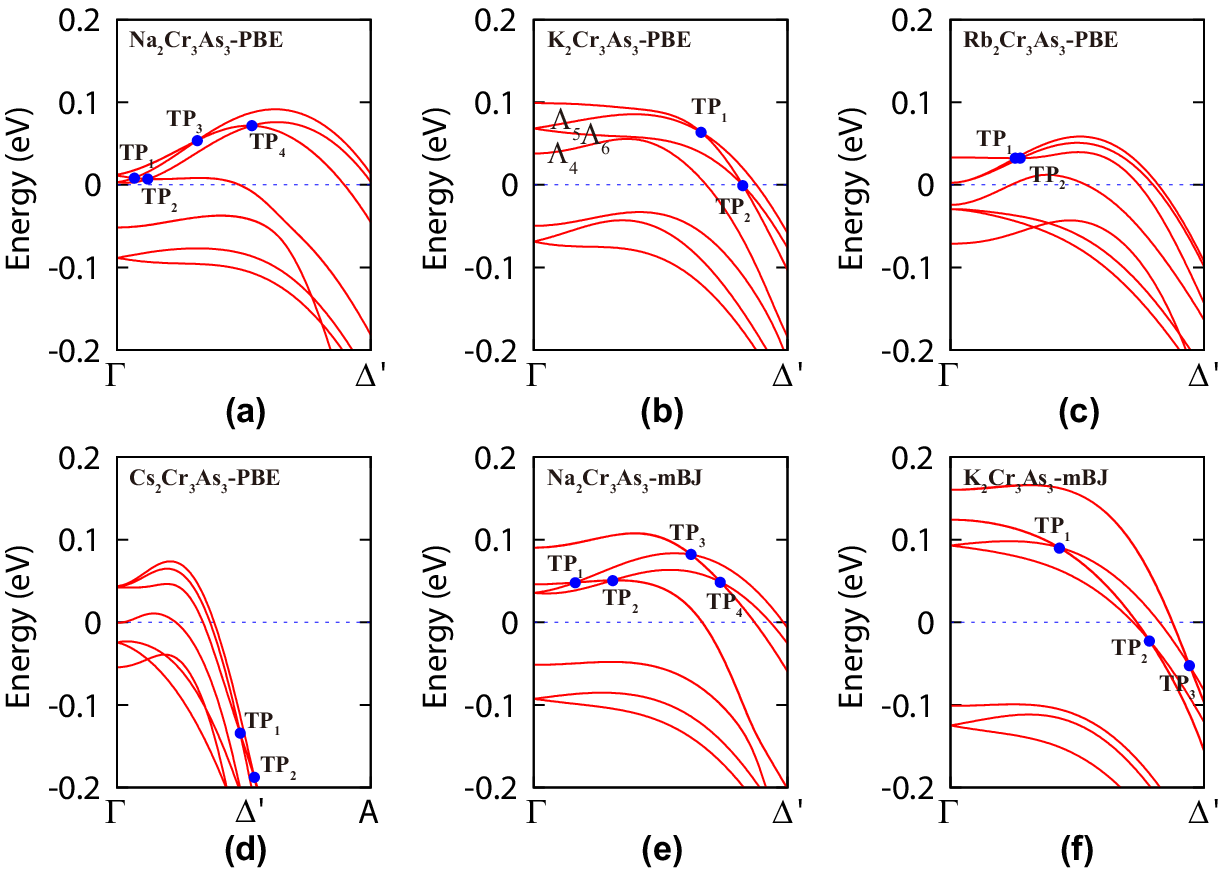}
  \caption{(a-d) The electronic structure of $A_2$Cr$_3$As$_3$ ($A$=Na,K,Rb and Cs) along $k_{z}$ within PBE calculations. (e-f) The electronic structures of Na$_2$Cr$_3$As$_3$ and K$_2$Cr$_3$As$_3$ within mBJ calculations.\label{fig:bs_all_kz}}
\end{figure}

\begin{figure}
  \includegraphics[width=12 cm]{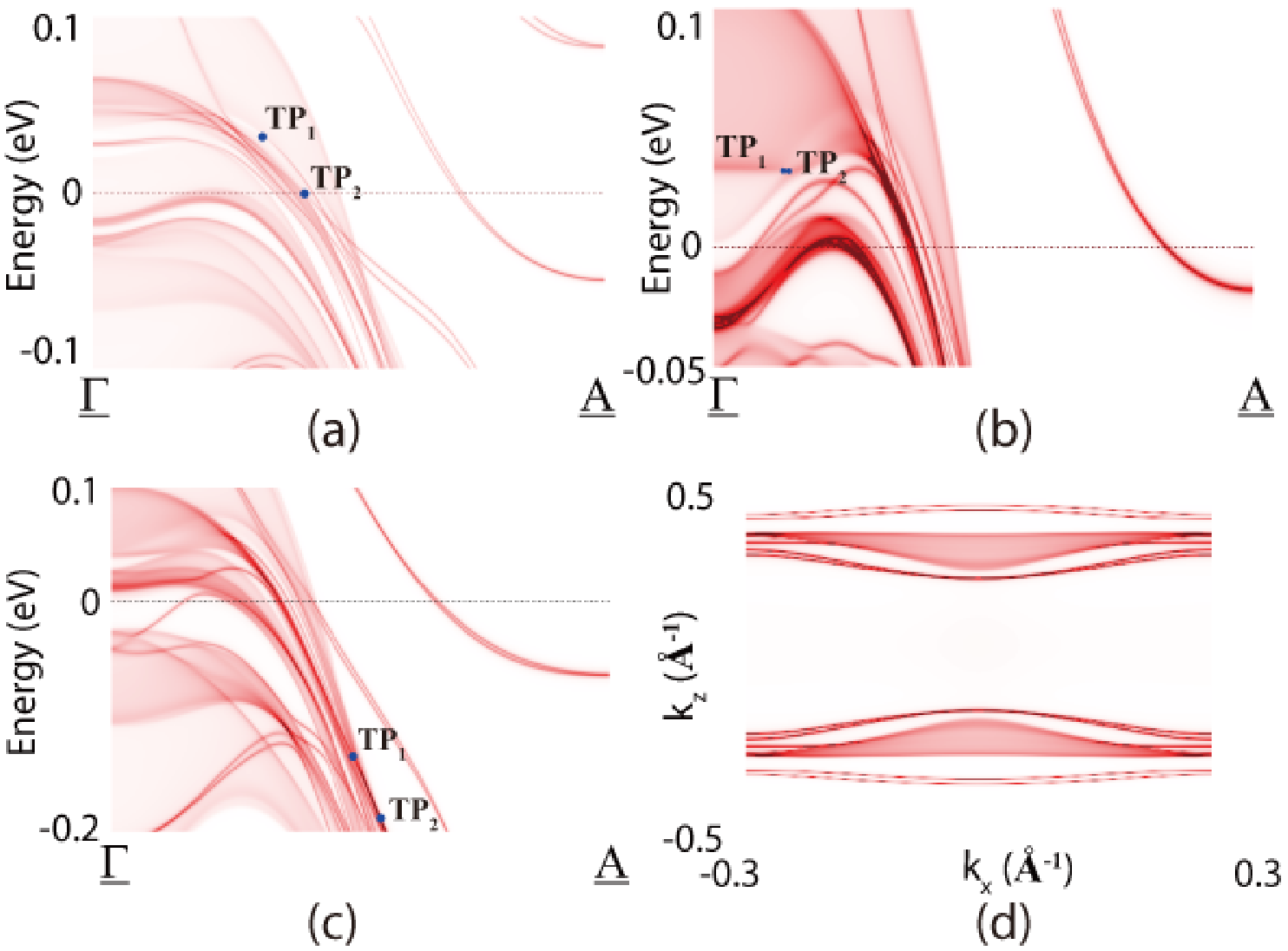}
 \caption{(a-c) The (010) surface state of $A_2$Cr$_3$As$_3$ ($A$=K,Rb and Cs) along $k_{z}$ with triplet points marked as blue dots within PBE calculations. (d) The iso-energy surface state at $\epsilon_F+0.15$ eV of K$_2$Cr$_3$As$_3$ within mBJ calculations
\label{fig:Ss_PBE}}
\end{figure}

\subsection{Multiorbital RPA Calculation}

\begin{figure}
 \includegraphics[width=10 cm]{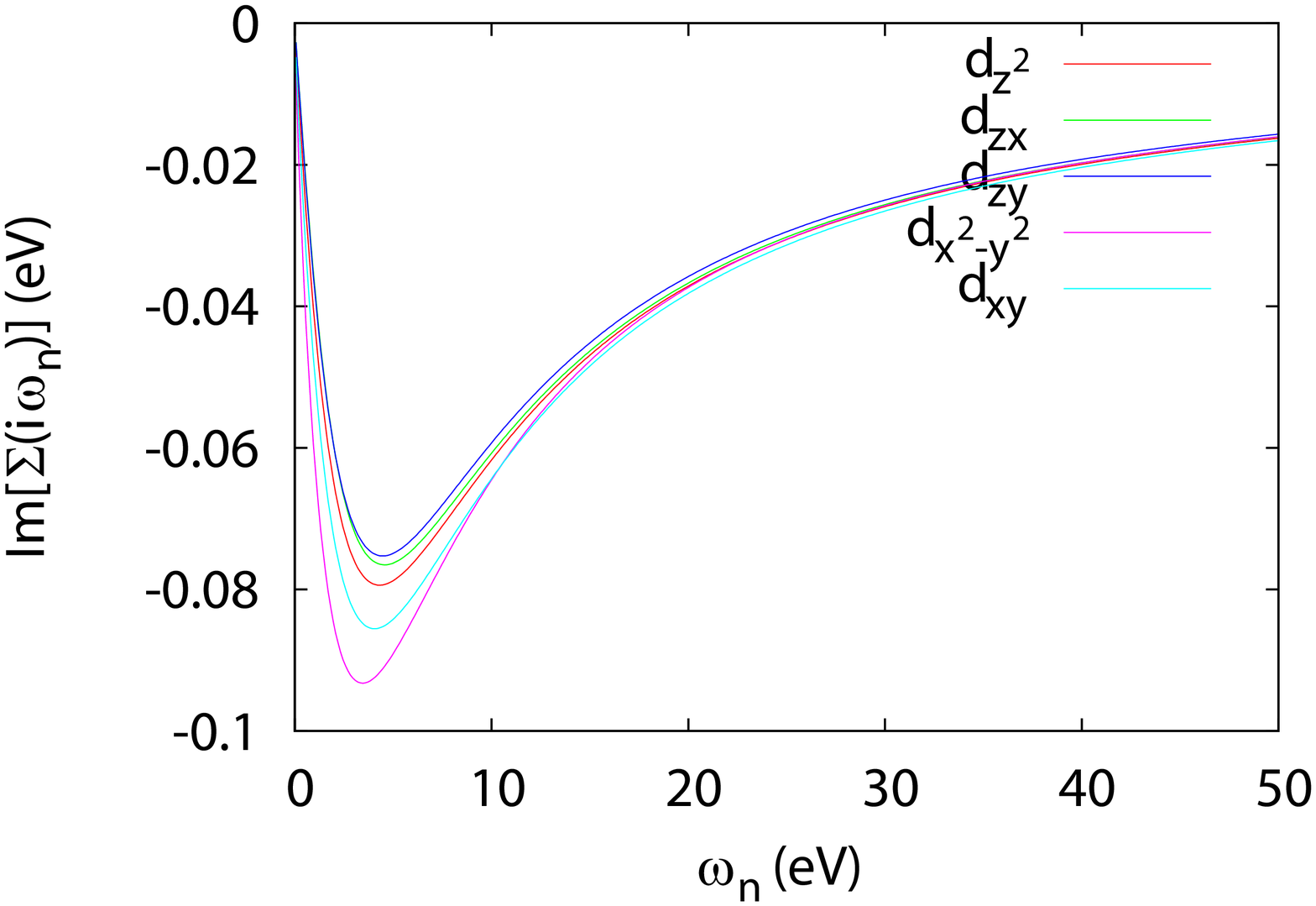}
  \caption{The Imaginary part of self-energy $\Sigma(i \omega, \mathbf{k})$ as a function of Matsubara frequency at
  $\mathbf{k}$=(0, 0, 0) point for Cr-$3d$ orbitals of K$_2$Cr$_3$As$_3$. A 3$\times$3$\times$6 $\Gamma$-centered q-point mesh was set to ensure the convergence of self-energy.
\label{fig:self-energy}}
\end{figure}

We show the calculated Matsubara frequency dependent self-energy $\Sigma(i \omega_n,\mathbf{k})$ (K$_2$Cr$_3$As$_3$) at $\Gamma$ point in Fig.~\ref{fig:self-energy}.

\subsection{Topological property in superconducting state}

\begin{table*}
\caption{List of occupation numbers $n_{\mathbf{k}}^{\alpha}$ associated with the six-fold rotoinversion symmetry $S^{z}_{6}$ eigenvalues $e^{i \frac{\pi\alpha}{6}}$($\alpha=\pm1,\pm3,\pm5$) at relevant k-points for K$_2$Cr$_3$As$_3$ mBJ results. The results for Na$_2$Cr$_3$As$_3$ (mBJ) are similar.
\label{tab:symme_occ}}
\begin{tabular}{c| cc|cc|cc|cc|cc|cc}
  \hline
 $\mathbf{k}$ &\multicolumn{2}{c }{$\Gamma$ (0, 0, 0)}
              &\multicolumn{2}{c }{$K$ ($\frac{\pi}{3}$, $\frac{\pi}{3}$, 0)}
              &\multicolumn{2}{c }{$K'$ ($\frac{-\pi}{3}$, $\frac{-\pi}{3}$, 0)}
              &\multicolumn{2}{c }{$A$ (0, 0, $\pi$)}
              &\multicolumn{2}{c }{$H$ ($\frac{\pi}{3}$, $\frac{\pi}{3}$, $\pi$)}
              &\multicolumn{2}{c }{$H'$ ($\frac{-\pi}{3}$, $\frac{-\pi}{3}$, $\pi$)}\\
 \hline
 PG.          &\multicolumn{2}{c }{$D_{3h}$}
              &\multicolumn{2}{c }{$C_{3h}$}
              &\multicolumn{2}{c }{$C_{3h}$}
              &\multicolumn{2}{c }{$D_{3h}$}
              &\multicolumn{2}{c }{$C_{3h}$}
              &\multicolumn{2}{c }{$C_{3h}$}\\
 \hline
 $e^{+i \frac{\pi}{6}}$ &\multicolumn{2}{c }{9}
                        &\multicolumn{2}{c }{9}
                        &\multicolumn{2}{c }{9}
                        &\multicolumn{2}{c }{10}
                        &\multicolumn{2}{c }{10}
                        &\multicolumn{2}{c }{10} \\
 $e^{-i \frac{\pi}{6}}$ &\multicolumn{2}{c }{9}
                        &\multicolumn{2}{c }{9}
                        &\multicolumn{2}{c }{9}
                        &\multicolumn{2}{c }{10}
                        &\multicolumn{2}{c }{10}
                        &\multicolumn{2}{c }{10} \\
 $e^{+i \frac{3\pi}{6}}$&\multicolumn{2}{c }{9}
                        &\multicolumn{2}{c }{10}
                        &\multicolumn{2}{c }{9}
                        &\multicolumn{2}{c }{10}
                        &\multicolumn{2}{c }{10}
                        &\multicolumn{2}{c }{10} \\
 $e^{-i \frac{3\pi}{6}}$&\multicolumn{2}{c }{9}
                        &\multicolumn{2}{c }{9}
                        &\multicolumn{2}{c }{10}
                        &\multicolumn{2}{c }{10}
                        &\multicolumn{2}{c }{10}
                        &\multicolumn{2}{c }{10} \\
 $e^{+i \frac{5\pi}{6}}$&\multicolumn{2}{c }{9}
                        &\multicolumn{2}{c }{9}
                        &\multicolumn{2}{c }{9}
                        &\multicolumn{2}{c }{10}
                        &\multicolumn{2}{c }{10}
                        &\multicolumn{2}{c }{10} \\
 $e^{-i \frac{5\pi}{6}}$&\multicolumn{2}{c }{9}
                        &\multicolumn{2}{c }{9}
                        &\multicolumn{2}{c }{9}
                        &\multicolumn{2}{c }{10}
                        &\multicolumn{2}{c }{10}
                        &\multicolumn{2}{c }{10} \\
\hline
\end{tabular}
\end{table*}

To identify the topological property in superconducting state, We use the symmetry indicators method recently proposed in Refs. \cite{Ono_symmetry_indicators,ono_arxiv_2019refined}. Since the system is invariant under the six-fold rotoinversion symmetry $S^{z}_{6}$, We consider the  superconducting gap symmetry belongs to $A''$ representation for $D_{3h}$ group. The topological property of this system can be characterized by $\mathbb{Z}_2 \times (\mathbb{Z}_6)^2$. These topological indexes are related to the occupation numbers  $n_{\mathbf{k}}^{\alpha}$ at $\Gamma$, $K$, $K'$, $A$, $H$ and $H'$ in the normal state (Tab.~\ref{tab:symme_occ}). The resulting topological indexes $\nu_{S_6}=0$, $z'_6=1$ and $z''_6=-1$. The nonzero 3D $\mathbb{Z}_6$ indexes imply the nontrivial topological property in the superconducting state. Our results are in agreement with Liu $et$ $al.$\cite{Liu_233}, but with an alternative approach.

\section{Supplementary Methods}
\subsection{RPA Susceptbility}
We first obtain the bare electronic susceptibility $\chi_{0}$  with Lindhard formula:
\begin{equation}\nonumber
  \chi_{0}=-\frac{1}{N_{\mathbf{k}}}\sum_{st}\sum_{\mu\nu\mathbf{k}}
  \frac{\langle s|\mu\mathbf{k}\rangle \langle \mu\mathbf{k}|t\rangle
        \langle t|\nu\mathbf{k+q}\rangle \langle \nu\mathbf{k+q}|s\rangle }{\omega+\varepsilon_{\nu\mathbf{k+q}}-\varepsilon_{\mu\mathbf{k}}+i0^{+}}(f(\varepsilon_{\nu\mathbf{k+q}})-f(\varepsilon_{\mu\mathbf{k}})),
\end{equation}
where $s$,$t$ are orbital indexes and $\mu$, $\nu$ are band indexes. Since the calculation is performed in the paramagnetic state, the spin index of above formula is omitted. We then consider the Hubbard-type Hamiltonian \cite{PhysRevB.75.224509,PhysRevB.69.104504}:
\begin{equation}
H=\sum\limits_{\mathbf{k}m\sigma}\varepsilon_{km\sigma}c^{\dagger}_{km\sigma}c_{km\sigma}
 +H_{int}, \nonumber
\end{equation}
where $H_{int}$ is the interaction part,
\begin{align}
H_{int}=U\sum_{is}n_{is\sigma}n_{is\overline{\sigma}}
  +U'\sum_{i,s,t \neq s}\sum_{\sigma,\sigma'}
  c^{\dagger}_{is\sigma}c^{\dagger}_{it\sigma'}c_{is\sigma'}c_{it\sigma} \nonumber \\
  +J\sum_{i,s,t \neq s}\sum_{\sigma,\sigma'}
  c^{\dagger}_{is\sigma}c^{\dagger}_{it\sigma'}c_{is\sigma'}c_{it\sigma}
  +J'\sum_{i,s,t \neq s}c^{\dagger}_{is\sigma}c^{\dagger}_{is\overline{\sigma}}
   c_{it \overline{\sigma}}c_{it\sigma'}  \nonumber
\end{align}
As mentioned in the main text, $U$, $U'$, $J$ and $J'$ denote the intra-orbital
Coulomb, inter-orbital Coulomb, Hund's coupling, and pair hopping interaction respectively. With the above multiorbital Hamiltonian,The charge and spin susceptibilities are obtained after the summation of bubble diagrams,
\begin{equation*}\nonumber
[\chi_{c}]_{pq;st}=\frac{[\chi_{0}]_{pq;st}}{I_{wz;pq}+[\chi_{0}]_{wz;uv}[U^{c}]_{uv;pq}}
\end{equation*}
and
\begin{equation*}\nonumber
[\chi_{s}]_{pq;st}=\frac{[\chi_{0}]_{pq;st}}{I_{wz;pq}-[\chi_{0}]_{wz;uv}[U^{s}]_{uv;pq}},
\end{equation*}
where $\chi_{0}$ is bare electronic susceptibility and $U^{c}$ ($U^{s}$) is interaction matrix of charge (spin) channel. The nonzero element of $U^{c}$ ($U^{s}$)\cite{Graser_2009,Graser_2010} are:
\begin{align}
[U^{c}]_{ss;ss}=U, [U^{c}]_{ss;tt}=2U'-J,
[U^{c}]_{st;st}=2J'-U',[U^{c}]_{st;ts}=J',  \nonumber
\end{align}
\begin{align}
[U^{s}]_{ss;ss}=U, \quad [U^{s}]_{ss;tt}=J,\qquad
[U^{s}]_{st;st}=U',\qquad  [U^{s}]_{st;ts}=J' \nonumber
\end{align}
Note that different from Wu $et$ $al.$\cite{Hu_K233} whose Hamiltonian is based on delocalized molecule orbitals with much weaker
on-site interactions, our Hamiltonian is constructed with maximally projected atomic Wannier functions, with $U$, $U'$, $J$, and $J'$ directly put on Cr-$3d$ orbitals.
The final charge and spin susceptibilities are obtained through:
\begin{equation}\nonumber
  \chi_{c(s)}(\mathbf{q},\omega)=\frac{1}{2}\sum_{st}[\chi_{c(s)}]_{ss,tt}(\mathbf{q},\omega)
\end{equation}

\subsection{RPA Self-Energy}
Within RPA, the single-particle Green's  function $G$ can be obtained using Dyson equation $G^{-1}(\mathbf{k}, \omega)=G_{0}^{-1}(\mathbf{k}, \omega)-\Sigma(\mathbf{k}, \omega)$. The self-energy $\Sigma$ is calculated in Matsubara frequency domain:
\begin{equation}\nonumber
  \Sigma(i \omega_n,\mathbf{k})=\sum_{i \nu_m, \mathbf{q}}G_0(i \omega_n-i \nu_m,\mathbf{k}-\mathbf{q})W(i \nu_m,\mathbf{q})
\end{equation}
where $W(i \nu_m,\mathbf{q})$ is the screened Coulomb interaction obtained from another Dyson-like equation $W^{-1}(i\nu_m, \mathbf{q})=v^{-1}-\chi(i\nu_m, \mathbf{q})$, and $v$ is the bare density-density type interaction (the $U$ and $U'$ terms in above equations). The real-frequency self-energy $\Sigma(\mathbf{k}, \omega)$ is then obtained by analytic continuation using P\`{a}de approximation.

\end{appendices}

\end{document}